\documentclass[notitlepage]{revtex4-2}
\usepackage{amsmath}
\usepackage{bm}
\usepackage{natbib}
\usepackage{graphicx}
\usepackage{epstopdf}
\usepackage{color}
\usepackage{soul}
\usepackage{hyperref}

\begin{document}
\title{High Weissenberg number simulations with incompressible Smoothed Particle Hydrodynamics and the log-conformation formulation}

\author{J. R. C. King}
\email{jack.king@manchester.ac.uk}
\affiliation{Department of Mechanical, Aerospace and Civil Engineering, The University of Manchester, UK}
\author{S. J. Lind}
\affiliation{Department of Mechanical, Aerospace and Civil Engineering, The University of Manchester, UK}


\date{\today}

\begin{abstract}

Viscoelastic flows occur widely, and numerical simulations of them are important for a range of industrial applications. Simulations of viscoelastic flows are more challenging than their Newtonian counterparts due to the presence of exponential gradients in polymeric stress fields, which can lead to catastrophic instabilities if not carefully handled. A key development to overcome this issue is the log-conformation formulation, which has been applied to a range of numerical methods, but not previously applied to Smoothed Particle Hydrodynamics (SPH). Here we present a 2D incompressible SPH algorithm for viscoelastic flows which, for the first time, incorporates a log-conformation formulation with an elasto-viscous stress splitting (EVSS) technique. The resulting scheme enables simulations of flows at high Weissenberg numbers (accurate up to $Wi=85$ for Poiseuille flow). The method is robust, and able to handle both internal and free-surface flows, and a range of linear and non-linear constitutive models. Several test cases are considerd included flow past a periodic array of cylinders and jet buckling. This presents a significant step change in capabilties compared to previous SPH algorithms for viscoelastic flows, and has the potential to simulate a wide range of new and challenging applications.

\end{abstract}

\maketitle

\section{Introduction}\label{intro}

Viscoelastic flows are widespread, both in the natural and man-made world. The applications of simulations of viscoelastic flows are numerous: from geophysical to biomedical to industrial and manufacturing processes. The behaviour of viscoelastic flows can be quite different to their Newtonian counterparts, exhibiting phenomena such as the Weissenberg effect, extrudate swelling and melt fracture~\cite{morozov_2007}, and elastic turbulence promoting mixing and increasing drag~\cite{groisman_2000}. Many of these phenomana can have important implications for industry. For example, melt fracture is the rate limiting factor in many extrusion processes~\cite{morozov_2007}. The motivation to better understand viscoelastic flows is clear, from both an intellectual and practical perspective, and numerical simulations are an important tool to that end.

The high Weissenberg number problem (HWNP)~\cite{keunings_1986} has long been \emph{the} major obstacle for computational rheology. The problem manifests as a catastrophic numerical instability which occurs at moderate Weissenberg numbers (typically $Wi\approx{1}$). The HWNP is caused by the nature of the the evolution equations for polymeric stresses: the velocity gradient acts multiplicatively on the stress, leading to exponential gradients in stress near boundaries and geometric singularities. Polynomial based reconstruction schemes cannot accurately capture the exponential gradients, underestimating fluxes and hence overestimating stress growth rates, resulting in numerical instability~\cite{fattal_2005}. 

The conformation tensor is a second order tensor which provides a macroscopic representation of the internal deformation of polymer chains, and is related to the polymeric stress via a strain function. A key advance in computational rheology was introduced by~\citet{fattal_2004,fattal_2005}, who recast the evolution equation for the conformation tensor in terms of its matrix-logarithm, an approach referred to as the log-conformation formulation. This has the effect of linearising the exponential gradients, enabling stable simulations with polynomial based schemes. The conformation tensor is symmetric positive definite (SPD) by definition. However, the SPD property is not necessarily conserved during numerical integration of the constitutive equations, and numerical results may consist of macroscopic states for which the corresponding molecular deformation is not physically valid. The log-conformation formulation ensures that the conformation tensor remains SPD by construction. A number of other transformations have been applied to stabilise the evolution equation of the conformation tensor, including the square-root transformation of~\cite{balci_2011}, and a generalised framework for kernel transformations of the conformation tensor~\cite{afonso_2012}. The log-conformation formulation has remained the most popular, being successfully applied in Finite Volume (FV)~\cite{lopez_2019}, Finite Element methods (FEM)~\cite{hulsen_2005} and Finite Difference (FD)~\cite{comminal_2015}, yielding stable simulations at Weissenberg numbers of $10$ and higher.

In this paper the log-conformation formulation is applied in the Smoothed Particle Hydrodynamics method for the first time. Smoothed Particle Hydrodynamics (SPH) is a mesh-free method, originally developed for astrophysical simulations~\cite{gingold_1977,lucy_1977}, since applied to a range of engineering problems, including compressible and incompressible flows, free surface flows, fluid structure interactions and solid mechanics~\cite{monaghan_2012}. Fluids are discretised by a set of Lagrangian particles, and spatial gradients are approximated by weighted sums of fluid properties at neighbouring particles. Free-surface flows pose difficulties for mesh-based methods, in which tracking a deforming surface undergoing topological changes is a complex task. Taking the example of simulations of buckling jets, conventional mesh-based approaches require highly resolved mesh in the region of the free surface (e.g.~\cite{bonito_2006}), or complex interface tracking algorithms(e.g.~\cite{tome_2002,tome_2019}). SPH is ideally suited to modelling free-surface and multiphase flows, as interfaces are handled naturally due its Lagrangian nature, and topological changes (e.g. jet coiling or bubble breakup) do not require any special treatment. 

The majority of SPH simulations of liquids follow a weakly compressible framework (WCSPH), where pressure is related to density via a stiff isothermal equation of state. Viscoelastic models have been implemented in WCSPH by several authors~\cite{ellero_2002,fang_2006,rafiee_2007,ren_2012,jiang_2012,xu_2013,vahabi_2019}. The schemes of~\cite{ellero_2002,fang_2006,rafiee_2007,ren_2012, jiang_2012,xu_2013} all require stabilisation with artificial viscosity or artificial stress. Work on viscoelastic SPH has also been published in the computer graphics community (e.g.~\cite{andrade_2015}), although in this field the focus is on the aesthetics of the results, rather than accurate representation of the physics. With the exception of~\cite{ellero_2002}, in the above schemes the density variation in the weakly compressible formulation is neglected in the constitutive models. Although SPH involves polynomial based reconstruction, and is therefore susceptible to the HWNP, all the above schemes integrate an evolution equation for the polymeric stress. We are only aware of one viscoelastic SPH formulation~\cite{vazquez_2009,vazquez_2012,grilli_2013,vazquez_2017}, based on the general equation for non-equilibrium reversible-irreversible coupling (GENERIC) formalism, in which the elastic stresses are calculated by evolution of the conformation tensor equation, and the authors acknowledge in~\cite{vazquez_2012} that positive definiteness of the conformation tensor is not ensured by construction. They postulate that this is due to the Lagrangian nature of SPH, which bypasses the non-linear convective terms in the evolution equation for the conformation tensor. In recent years, incompressible SPH (ISPH) has gained popularity, and is the SPH variant employed herein. Originally developed by~\cite{cummins_1999}, and based on the projection method of~\citet{chorin_1968}, incompressibility is enforced by solving a Poisson equation for pressure. 

It is of interest at this point to compare the GENERIC-based approach of~\cite{vazquez_2009} with ISPH. The GENERIC-based approach bears similarities (qualitatively) to the Hamiltonian view of SPH common in astrophysics~\cite{price_2012}. In both~\cite{vazquez_2009} and~\cite{price_2012}, the focus is on global conservation, and the symmetries which give rise to the conservation laws of interest are retained in the SPH discretisation at the level of particle-particle interactions. With the work of~\cite{vazquez_2009,vazquez_2017}, the resulting discretisation ensures that the first and second laws of thermodynamics, and the fluctuation-dissipation theorem are respected. The resulting discrete system of equations can be shown to correspond to the continuum equations (in~\cite{vazquez_2009}, including the form of the constitutive law), despite the continuum equations playing no role in the derivation of the scheme. This is in contrast to ISPH, where the numerical scheme is constructed by applying discrete SPH derivative operators to the continuum equations. In ISPH consistent (and non-conservative) operators are popular (e.g.~\cite{lind_2012,zainali_2013}), and global conservation is tied to resolution and consistency. In this regard, ISPH may be viewed as an approximate finite difference scheme. In incompressible Lagrangian SPH with consistent gradients, particles more accurately follow streamlines, leading to severe anisotropy of particle distributions, especially around stagnation points. The development of Fickian shifting (also called ``particle shifting'') by~\cite{xu_2009,lind_2012,skillen_2013} provided stabilisation of ISPH, substantially increasing the practical value of the method. Whilst particle shifting can disrupt exact global conservation, this is of no detriment in many ISPH schemes, where such symmetries are already broken by use of consistent operators. Whilst ISPH has been adapted to study generalised Newtonian fluids~\cite{xenakis_2015}, the only implementation of viscoelasticity in ISPH (of which the authors are aware) is that of~\cite{zainali_2013}. Note, whilst typical SPH simulations exhibit convergence rates of between $1$ and $2$, recently an Eulerian variant of SPH has been proposed, with high order convergence demonstrated~\cite{lind_2016}.

In this paper we present a new numerical scheme for simulations of incompressible viscoelastic fluids with SPH. The Navier Stokes equations are solved in an incompressible Arbitrary Lagrangian-Eulerian (ALE) framework, and constitutive equations are integrated using the log-conformation approach combined with an elasto-viscous stress splitting (EVSS) scheme. The EVSS technique, originally developed for finite element simulations~\cite{beris_1984,rajagopalan_1990}, is now widely used for simulations of viscoelastic flows, particularly with low or zero solvent viscosities, in a range of numerical methods (see e.g.~\cite{tome_2019} for a finite difference implementation).  To the best of our knowledge, this is the first implementation of either the log-conformation formulation or the EVSS scheme in a mesh-free method. We introduce improved free surface boundary conditions for SPH, and a modification to the shifting algorithm which provides more accurate boundary representation for free-surface low Reynolds number flows. This allows robust simulations of internal and free surface flows at significantly higher Weissenberg (and Deborah) numbers than possible in previous SPH schemes. The remainder of this paper is set out as follows. In Section~\ref{ge} we present the governing equations and the log-conformation formulation. In Section~\ref{nummeth} we detail the numerical method. In Section~\ref{numres} we present results for a range of test internal and free surface flow problems, providing validation against analytical and published numerical results. Section~\ref{conc} is a summary of our conclusions.

\section{Governing equations}\label{ge}

We consider the two-dimensional incompressible Navier-Stokes (NS) equations, which may be written in an arbitrary frame of reference as:
\begin{equation}\nabla\cdot\bm{u}=0,\label{eq:mass}\end{equation}
\begin{equation}\frac{d\bm{u}}{dt}-\bm{u_{ps}}\cdot\nabla\bm{u}=-\frac{1}{\rho}\nabla p+\frac{1}{\rho}\nabla\cdot\bm{\tau}+\bm{g},\label{eq:mom}\end{equation}
where $p$ is the pressure, $\bm{\tau}$ is the extra stress, $\bm{u}=\left[u,v\right]^{T}$ is the velocity of the fluid, $\rho$ is the density and $\bm{g}$ is a body force. The derivative $d\left(\cdot\right)/dt$ is the time derivative \emph{in the local arbitrary frame of reference}. The vector field $\bm{u_{ps}}$ is the difference between the velocity of the local frame of reference, and the velocity of the fluid. If we set $\bm{u_{ps}}=\bm{0}$, then we obtain in~\eqref{eq:mom} the NS equations in a Lagrangian frame of reference. If $\bm{u_{ps}}=-\bm{u}$,  we obtain the Eulerian form of the NS equations. This ALE framework allows us apply the shifting technique of~\cite{lind_2012} in a computationally efficient manner, setting $\bm{u_{ps}}$ according to a Fickian diffusion equation, which improves SPH particle distributions and numerical stability. Although in this work we consider only the cases of Eulerian and (quasi-)Lagrangian SPH, the formulation presented here may be applied to any arbitrary frame of reference. The extra stress is decomposed into solvent and polymeric components:
\begin{equation}\bm{\tau}=\bm{\tau_{s}}+\bm{\tau_{p}}.\end{equation}
The solvent stress is purely viscous and, defining the deformation rate tensor $\bm{D}=\left(\nabla\bm{u}+\nabla\bm{u}^{T}\right)/2$, is given by
\begin{equation}\bm{\tau_{s}}=2\eta_{s}\bm{D}\end{equation}
where $\eta_{s}$ is the solvent viscosity, $\eta_{0}$ is the total viscosity, and $\beta$ is the viscosity ratio, defined as $\beta=\eta_{s}/\eta_{0}$. Hence the viscous term in~\eqref{eq:mom} due to the solvent viscosity is
\begin{equation}\frac{1}{\rho}\nabla\cdot\bm{\tau_{s}}=\frac{\eta_{s}}{\rho}\nabla^{2}\bm{u}.\end{equation}
In this work we employ an EVSS technique, for which we introduce the tensor $\bm{\Phi}$ as
\begin{equation}\bm{\Phi}=\bm{\tau_{p}}-\alpha_{V}\eta_{0}2\bm{D}\label{eq:evss}.\end{equation}
The extra stress $\bm{\tau}$ is then
\begin{equation}\bm{\tau}=\bm{\Phi}+\left(\beta+\alpha_{V}\right)\eta_{0}2\bm{D},\end{equation}
and the governing equation~\eqref{eq:mom} may be written
\begin{equation}\frac{d\bm{u}}{dt}-\bm{u_{ps}}\cdot\nabla\bm{u}=-\frac{1}{\rho}\nabla p+\frac{1}{\rho}\nabla\cdot\bm{\Phi}+\frac{\left(\beta+\alpha_{V}\right)\eta_{0}}{\rho}\nabla^{2}\bm{u}+\bm{g}.\label{eq:mom_evss}\end{equation}
This formulation introduces some viscosity into the numerical scheme, allowing stable simulations even when $\beta=0$. The parameter $\alpha_{V}$ controls the amount of viscosity, with $\alpha_{V}=0$ being no additional viscosity (equivalent to not using the EVSS scheme) and $\alpha_{V}=1-\beta$ being the standard EVSS scheme (e.g.~\cite{tome_2019}). Note that we do not adapt $\alpha_{V}$ either spatially or temporally (as done by some authors e.g.~\cite{sun_1996}). The chosen value of $\alpha_{V}$ is case dependent and used only where necessary (and explicitly mentioned) for stability. 

The polymeric stress is related to the conformation tensor $\bm{A}$ via the strain function
\begin{equation}\bm{\tau_{p}}=\frac{\eta_{p}}{\lambda}f_{S}\left(\bm{A}\right),\label{eq:tau_A}\end{equation}
where the polymeric viscosity $\eta_{p}=\eta_{0}-\eta_{s}=\left(1-\beta\right)\eta_{0}$, $\lambda$ is the relaxation time, and the specific form of the strain function $f_{S}$ is determined by the choice of constitutive equation. The conformation tensor $\bm{A}$ is a macroscropic representation of the average orientation of the polymer chains, and from this definition is symmetric positive definite (SPD). Where deformation is zero, $\bm{A}=\bm{I}$, the identity matrix. The conformation tensor obeys an evolution equation
\begin{equation}\overset{\nabla}{\bm{A}}=-\frac{1}{\lambda}f_{R}\left(\bm{A}\right),\label{eq:const_eq}\end{equation}
in which $\overset{\nabla}{\bm{A}}$ is the upper-convected derivative of $\bm{A}$, defined by
\begin{equation}\overset{\nabla}{\bm{A}}=\frac{d\bm{A}}{dt}-\bm{u_{ps}}\cdot\nabla\bm{A}-\left(\bm{A}\cdot\nabla\bm{u}^{T}+\nabla\bm{u}\cdot\bm{A}\right),\end{equation}
and $f_{R}$ is a relaxation function, specific to the choice of constitutive equation. Equation~\eqref{eq:const_eq} describes the response and relaxation of the polymer deformation to the macroscopic deformation. For many popular constitutive models the strain and relaxation functions are polynomials of $\bm{A}$, usually first or second order, with coefficients which may depend of the invariants of $\bm{A}$. In this paper, we consider a number of constitutive models: the Oldroyd B, Upper Convected Maxwell (UCM), Finite Extensible Non-linear Elastic (FENE), Phan-Thien-Tanner (PTT) and Giesekus models. The strain and relaxation functions of these are detailed in Table~\ref{tab:cm}. Although the linear-elasticity of the Oldroyd B model can lead to localised negative viscosities at certain shear rates, it is a convenient model for benchmarking numerical schemes, because analytical solutions exist for a number of flow geometries (e.g. unsteady channel flow). Furthermore, the non-linearity present in other models has a stabilising effect absent from the Oldroyd B model, making it a useful model with which to test the limits of stability of a numerical scheme. The UCM model is the limiting case of the Oldroyd B model with $\beta=0$. The FENE, PTT, and Giesekus models include non-linear elasticity via the parameters $L^{2}$, $\varepsilon$ and $\alpha$ respectively. In the limits of $L^{2}\to\infty$, $\varepsilon=0$ and $\alpha=0$, these models collapse to the Oldroyd B model. To avoid non-physical solutions for the Giesekus model, there is an upper limit of $\alpha\le0.5$.

\begin{table}
\begin{center}
\caption{Strain and relaxation functions for the constitutive models used in this work.\label{tab:cm}}
\begin{tabular}{|l|l|l|}
Constitutive model & $f_{S}$& $f_{R}$\\
\hline
Oldroyd B & $\bm{A}-\bm{I}$ & $\bm{A}-\bm{I}$ \\
FENE-P & $\frac{\bm{A}}{1-tr\left(\bm{A}\right)/L^{2}}-\bm{I}$ & $\frac{\bm{A}}{1-tr\left(\bm{A}\right)/L^{2}}-\bm{I}$ \\
FENE-CR & $\frac{\bm{A}-\bm{I}}{1-tr\left(\bm{A}\right)/L^{2}}$ & $\frac{\bm{A}-\bm{I}}{1-tr\left(\bm{A}\right)/L^{2}}$ \\
Linear PTT & $\bm{A}-\bm{I}$ & $\left[1+\varepsilon{tr}\left(\bm{A}-\bm{I}\right)\right]\left(\bm{A}-\bm{I}\right)$ \\
Exponential PTT & $\bm{A}-\bm{I}$ & $\exp\left[\varepsilon{tr}\left(\bm{A}-\bm{I}\right)\right]\left(\bm{A}-\bm{I}\right)$ \\
Giesekus & $\bm{A}-\bm{I}$ & $\alpha\bm{A}^{2}+\left(1-2\alpha\right)\bm{A}-\left(1-\alpha\right)\bm{I}$ \\
\end{tabular}
\end{center}
\end{table}

\subsection{The Log-conformation formulation}
To overcome the instabilities encounted when integrating $\bm{A}$, and to ensure $\bm{A}$ remains SPD, we use the log-conformation formulation introduced by~\citet{fattal_2004,fattal_2005}. We denote the logarithm of the conformation tensor $\bm{\Psi}=\log\bm{A}$. Given that $\bm{A}$ is symmetric positive-definite, it can be expressed in terms of its eigenvectors and eigenvalues as:
\begin{equation}\bm{A}=\bm{R}\bm{\Lambda}\bm{R}^{T}\label{eq:diag},\end{equation}
in which $\bm{R}$ is a matrix of the eigenvectors of $\bm{A}$, and $\bm{\Lambda}$ is a diagonal matrix of the eigenvalues ($\Lambda_{1}$ and $\Lambda_{2}$ in two-dimensions). Combining the expression for $\bm{\Psi}$ with~\eqref{eq:diag} we obtain
\begin{equation}\bm{\Psi}=\bm{R}\ln\bm{\Lambda}\bm{R}^{T}\label{eq:Psi},\end{equation}
where the logarithm is applied element-wise to the non-zero elements of $\bm{\Lambda}$. The velocity gradient is decomposed as
\begin{equation}\nabla\bm{u}=\bm{\Omega}+\bm{B}+\bm{NA}^{-1},\label{eq:diag_gu}\end{equation}
where $\bm{\Omega}$ and $\bm{N}$ are antisymmetric, and $\bm{B}$ is symmetric and traceless. Substituting~\eqref{eq:Psi} and~\eqref{eq:diag_gu} into~\eqref{eq:const_eq} we obtain an evolution equation for $\bm{\Psi}$:
\begin{equation}\frac{d\bm{\Psi}}{dt}-\bm{u_{ps}}\cdot\nabla\bm{\Psi}-\left(\bm{\Omega\Psi}-\bm{\Psi\Omega}\right)-2\bm{B}=-\frac{1}{\lambda}\exp\left(-\bm{\Psi}\right)f_{R}\left(\exp\left(\bm{\Psi}\right)\right)\label{eq:logconf_evol}\end{equation}
The components of the decomposed velocity gradient are given by
\begin{equation}\begin{bmatrix}m_{11}&m_{12}\\m_{21}&m_{22}\end{bmatrix}=\bm{R}^{T}\nabla\bm{u}\bm{R}\label{eq:dc_gu1}\end{equation}
\begin{equation}\bm{B}=\bm{R}\begin{bmatrix}m_{11}&0\\0&m_{22}\end{bmatrix}\bm{R}^{T}\end{equation}
\begin{equation}\bm{\Omega}=\bm{R}\begin{bmatrix}0&\omega\\-\omega&0\end{bmatrix}\bm{R}^{T},\end{equation}
with $\omega$ defined by
\begin{equation}\omega=\frac{\Lambda_{2}m_{12}+\Lambda_{1}m_{21}}{\Lambda_{2}-\Lambda_{1}}.\label{eq:dc_gu2}\end{equation}

\section{Numerical method}\label{nummeth}

In SPH, the fluid is represented by set of discrete particles, each of which we label $i\in\left[1,N\right]$, where $N$ is the total number of particles. Each particle $i$ has a volume $\delta{V}$ (uniform and constant in the present formulation) and a position vector $\bm{r}_{i}=\left[x_{i},y_{i}\right]^{T}$. The velocity, polymeric stress and pressure of each particle $i$ are denoted $\bm{u}_{i}=\left[u_{i},v_{i}\right]^{T}$, $\bm{\tau}_{i}$, and $p_{i}$ respectively. We denote the difference in the property $\left(\cdot\right)$ of two particles $i$ and $j$ as $\left(\cdot\right)_{ij}=\left(\cdot\right)_{i}-\left(\cdot\right)_{j}=-\left(\cdot\right)_{ji}$. In SPH, gradients of field variables at the location of particle $i$ are calculated using a weighted sum of the values of the field variables at the neighbouring particles $j\in\mathcal{N}_{i}$ where the weights are obtained from a kernel function $W\left(\left\lvert\bm{r}_{ij}\right\rvert\right)=W_{ij}$ and its derivatives. The set of neighbours $\mathcal{N}_{i}$ contains all particles $j$ with $\left\lvert\bm{r}_{ij}\right\rvert\le{r}_{ks}$, where $r_{ks}$ is the support radius of the kernel. For a derivation and analysis of SPH fundamentals, we refer the reader to~\cite{price_2012,fatehi_2011,monaghan_2012}. Throughout this work we use a quintic-spline kernel~\cite{morris_1997} with a support radius of $3h$ and an initial particle spacing of $\delta{r}=h/1.3$, yielding stencils with approximately $44$ neighbours in regions of full kernel support. First derivatives are discretized according to
\begin{equation}\langle\nabla\phi\rangle_{i}=\displaystyle\sum_{j}\left(\phi_{j}-\phi_{i}\right)\nabla\hat{W}_{ij}\delta{V}\qquad\langle\nabla\cdot\bm{u}\rangle_{i}=\displaystyle\sum_{j}\left(\bm{u}_{j}\pm\bm{u}_{i}\right)\cdot\nabla\hat{W}_{ij}\delta{V}\label{eq:grad},\end{equation}
with the choice of positive or negative formulation of the divergence operator case dependent: when calculating $\nabla\cdot\bm{u}$ we use the negative version. When calculating $\nabla\cdot\bm{\Psi}$, the choice of sign in~\eqref{eq:grad} is determined by the flow being simulated. If free surfaces are present, we use the positive version which, although less accurate, better approximates the free-surface boundary condition~\cite{fang_2006}. For closed domains, we use the negative version, which is more accurate, and provides increased stability when large stress gradients are present. The Laplacian is approximated using the formulation of~\cite{morris_1997} as
\begin{equation}\langle\nabla^{2}\phi\rangle_{i}=\displaystyle\sum_{j}\frac{2\phi_{ij}}{\lvert\bm{r_{ij}}\rvert^{2}}\bm{r_{ij}}\cdot\nabla\hat{W}_{ij}\delta{V}.\label{eq:lap}\end{equation}
The corrected kernel gradient $\nabla\hat{W}_{ij}$ (note the circumflex) due to~\citet{bonet_lok} is used, which provides first-order consistency for first derivatives, and ensures zero-order consistency for the Laplacian, and is given by $\nabla\hat{W}_{ij}=\bm{L}_{i}\nabla{W}_{ij}$, with $\nabla{W}$ the uncorrected kernel gradient, and the correction tensor
\begin{equation}\bm{L}_{i}=\begin{bmatrix}\displaystyle\sum{x}_{ji}\nabla{W}_{ij}\cdot\bm{e_{x}}\delta{V}&\displaystyle\sum{x}_{ji}\nabla{W}_{ij}\cdot\bm{e_{y}}\delta{V}\\\displaystyle\sum{y}_{ji}\nabla{W}_{ij}\cdot\bm{e_{x}}\delta{V}&\displaystyle\sum{y}_{ji}\nabla{W}_{ij}\cdot\bm{e_{y}}\delta{V}\end{bmatrix}^{-1},\end{equation}
with $\bm{e_{x}}$ and $\bm{e_{y}}$ the unit vectors in the $x$- and $y$-directions respectively. When the positive version of~{\eqref{eq:grad}} is used, we set $\bm{L}_{i}=\bm{I}$, the identity tensor. 

\subsection{Incompressible SPH algorithm}
The system given by~\eqref{eq:mass},~\eqref{eq:mom}, and~\eqref{eq:const_eq} is solved using Chorin's projection method~\cite{chorin_1968}, initially introduced to SPH by~\cite{cummins_1999}. Velocities and pressures are stored at each time-step $n$, and the polymeric stress is calculated and stored at each half time-step $n+1/2$. The algorithm is presented in an Arbitrary-Lagrangian-Eulerian (ALE) framework, in which particles move with velocity $\bm{u}+\bm{u_{ps}}$. In the present work, we investigate the cases of quasi-Lagrangian (with a small $\bm{u_{ps}}$ providing numerical stability), and Eulerian schemes. For fully arbitrary $\bm{u_{ps}}$ the algorithm is unchanged, apart from the prescription of $\bm{u_{ps}}$. In the following, each operation is applied to every particle $i\in\left[1,N\right]$, and we have dropped the subscripts for clarity. The algorithm is as follows.
\begin{itemize}
\item\emph{Step 1:} Particles are advected to an intermediate position with velocity $\bm{u}^{n}$:
  \begin{equation}\bm{r}^{\star}=\bm{r}^{n}+\delta{t}\bm{u}^{n}\end{equation}
\item\emph{Step 2:} Mirror particles are created, with the appropriate properties (see Section~\ref{sec:bcs}), and neighbour lists $\mathcal{N}$ are built. For Eulerian SPH we set $\bm{u}_{s}=-\bm{u}$. For Lagrangian SPH, we calculate $\bm{u}_{s}$ based on a modified form of the Fickian shifting introduced by~\cite{lind_2012}, detailed in Section~\ref{sec:shift}.
\item\emph{Step 3:} The constitutive equation~\eqref{eq:const_eq} is integrated to obtain $\bm{\tau_{p}}^{n+1/2}$, as detailed in Section~\ref{sec:taup}, from which $\bm{\Phi}^{n+1/2}$ is calculated via~\eqref{eq:evss} 
\item\emph{Step 4:} An intermediate velocity $\bm{u}^{\star}$ is calculated by neglecting the pressure gradient in~\eqref{eq:mom}, which is included later, and integrating forward in time with according to
\begin{equation}\bm{u}^{\star}=\bm{u}^{n}+\delta{t}\left[\bm{u_{ps}}\cdot\nabla\bm{u}^{n}+\frac{\left(\beta+\alpha_{V}\right)\eta_{0}}{\rho}\nabla^{2}\bm{u}^{n}+\frac{1}{\rho}\nabla\cdot\bm{\Phi}^{n+1/2}+\bm{g}\right],\end{equation}
where $\nabla\cdot\bm{\Phi}^{n+1/2}$ is calculated using the positive version of~\eqref{eq:grad}. Note that previous versions of incompressible SPH (e.g.~\cite{lind_2012,chow_2018}) omit the advective term $\bm{u_{ps}}\cdot\nabla\bm{u}$ from this step, and instead shift particles with velocity $\bm{u_{ps}}$ at the end of each time step, once particle positions and properties have been updated. By including the advective term, we ensure that the resulting velocity field is divergence free (to first order in time). This approach has the benefit of reducing the number of times neighbour lists and velocity gradients must be calculated each time step (from twice to once), reducing computational costs.
\item\emph{Step 5:} The pressure at time $n+1$ is obtained from the pressure Poisson equation (PPE), written as
\begin{equation}\nabla\cdot\left(\frac{1}{\rho}\nabla{p}^{n+1}\right)=\frac{1}{\delta{t}}\nabla\cdot\bm{u}^{\star}.\end{equation}
The PPE, subject to appropriate boundary conditions as described in Section~\ref{sec:bcs}, is formulated as a sparse linear system, and solved using a BiCGStab algorithm with Jacobi preconditioning.
\item\emph{Step 6:} Next, the divergence free velocity field $\bm{u}^{n+1}$ is obtained from the projection of $\bm{u}^{\star}$:
\begin{equation}\bm{u}^{n+1}=\bm{u}^{\star}-\frac{1}{\rho}\nabla{p}^{n+1},\end{equation}
where $\nabla{p}$ is found using~\eqref{eq:grad}.
\item\emph{Step 7:} Finally, the time step is completed as the particles are advected to their final positions
\begin{equation}\bm{r}^{n+1}=\bm{r}^{n}+\frac{\delta{t}}{2}\left(\bm{u}^{n}+\bm{u}^{n+1}+2\bm{u_{ps}}^{n}\right).\end{equation}
This advection step combines the standard incompressible SPH advection step - using the average of the velocity field at time-steps $n$ and $n+1$ (see e.g.~\cite{lind_2012}), with a first order (in time) integration using the shifting velocity $\bm{u_{ps}}$.
\end{itemize}

In the specific case of Eulerian SPH, we omit steps 1 and 7. The value of $\delta{t}$ is set adaptively according to criteria for the Courant condition and viscous diffusion, as in~\cite{xenakis_2015}:
\begin{equation}\delta{t}=0.2\min\left(\frac{h}{\max_{i}\left(\lvert\bm{u}\rvert\right)},\frac{\rho{h}^{2}}{\eta_{0}}\right),\end{equation}
in which $\max_{i}$ is the maximum value over all particles $i\in\left[1,N\right]$.

\subsection{Integration of the polymeric stress}\label{sec:taup}

We store as primary variables the polymeric stress and the trace of the conformation tensor $tr\left(\bm{A}\right)$. At each time-step, the conformation tensor is calculated from $\bm{\tau_{p}}$ and $tr\left(\bm{A}\right)$. We then integrate the conformation tensor using a time-splitting procedure~\cite{hao_2007,lopez_2019}, before calculating and storing the new values of $\bm{\tau_{p}}$ and $tr\left(\bm{A}\right)$. The conformation tensor evolution equation~\eqref{eq:const_eq} is split between the advection and source terms, giving
\begin{subequations}
\begin{align}
\frac{d\bm{A}}{dt}&=\bm{u_{ps}}\cdot\nabla\bm{A}+\left(\bm{A}\cdot\nabla\bm{u}^{T}+\nabla\bm{u}\cdot\bm{A}\right)\label{eq:A_adv}\\
\frac{d\bm{A}}{dt}&=-\frac{1}{\lambda}f_{R}\left(\bm{A}\right)\label{eq:A_src}.
\end{align}
\end{subequations}
The same procedure applied to the equation for the evolution of $\bm{\Psi}$ yields
\begin{subequations}
\begin{align}
\frac{d\bm{\Psi}}{dt}&=\bm{u_{ps}}\cdot\nabla\bm{\Psi}+\left(\bm{\Omega\Psi}-\bm{\Psi\Omega}\right)+2\bm{B}\label{eq:psi_adv}\\
\frac{d\bm{\Psi}}{dt}&=-\frac{1}{\lambda}\exp\left(-\bm{\Psi}\right)f_{R}\left(\exp\left(\bm{\Psi}\right)\right).\label{eq:psi_src}
\end{align}
\end{subequations}
In our algorithm, we start with $\bm{\tau_{p}}^{n-1/2}$, and calculate $\bm{A}^{n-1/2}$. We then use the logarithm transformation to obtain $\bm{\Psi}^{n-1/2}$, and integrate~\eqref{eq:psi_adv} to obtain an intermediate $\bm{\Psi}^{\star}$. We then take the inverse logarithm transformation to obtain $\bm{A}^{\star}$, and then use~\eqref{eq:A_src} to obtain the new value of $\bm{A}^{n+1/2}$, from which $\bm{\tau_{p}}^{n+1/2}$ is calculated. 
The algorithm is as follows:
\begin{itemize}
\item\emph{Step 1:} Obtain the conformation tensor from $\bm{\tau_{p}}^{n-1/2}$ by inverting $f_{S}$ in~\eqref{eq:tau_A}. For constitutive equations where $f_{S}$ depends on $tr\left(\bm{A}\right)$, we use $tr\left(\bm{A}^{n-1/2}\right)$.
\item\emph{Step 2:} Diagonalise the conformation tensor, obtaining the eigenvectors and eigenvalues $\bm{R}$ and $\bm{\Lambda}$. Then calculate the log-conformation tensor $\bm{\Psi}^{n-1/2}$ from~\eqref{eq:Psi}. 
\item\emph{Step 3:} Calculate the velocity gradient $\nabla\bm{u}^{n}$ using~\eqref{eq:grad}, and decompose this to obtain $\bm{B}^{n}$ and $\bm{\Omega}^{n}$ from~\eqref{eq:dc_gu1} to~\eqref{eq:dc_gu2}.
\item\emph{Step 4:} Calculate $\nabla\bm{\Psi}^{n-1/2}$ using~\eqref{eq:grad}.
\item\emph{Step 5:} Having determined all the terms on the RHS of~\eqref{eq:psi_adv}, calculate an intermediate $\bm{\Psi}^{\star}$ using
\begin{equation}\bm{\Psi}^{\star}=\bm{\Psi}^{n-1/2}+\delta{t}\left[\bm{u_{ps}}^{n}\cdot\nabla\bm{\Psi}^{n-1/2}+\left(\bm{\Omega}^{n}\bm{\Psi}^{n-1/2}-\bm{\Psi}^{n-1/2}\bm{\Omega}^{n}\right)-2\bm{B}^{n}\right].\end{equation}
\item\emph{Step 6:} Diagonalise the log-conformation tensor $\bm{\Psi}^{\star}$, obtaining the eigenvectors and eigenvalues $\bm{R}$ and $\log\bm{\Lambda}$. Then calculate the conformation tensor $\bm{A}^{\star}=\bm{R\Lambda{R}}^{T}$.
\item\emph{Step 7:} How we integrate the source terms~\eqref{eq:A_src} depends on the constitutive equation. For the Oldroyd B, UCM, FENE and PTT models, we integrate~\eqref{eq:A_src} quasi-analytically following~\cite{lopez_2019}, and neglecting the change in $tr\left(\bm{A}\right)$ over a time step. For these constitutive models, we can write $f_{R}=\alpha_{R}\left(\beta_{R}\bm{A}-\bm{I}\right)$, with $\alpha_{R}$ and $\beta_{R}$ constants, or functions of $tr\left(\bm{A}\right)$, with value and form dependent on the consitutive model. Assuming $\alpha_{R}$ and $\beta_{R}$ are constant over the period of integration, we can integrate~\eqref{eq:A_src} analytically to find $\bm{A}^{n+1/2}$:
\begin{equation}\bm{A}^{n+1/2}=\bm{A}^{\star}\exp\left(\frac{-\alpha_{R}\beta_{R}\delta{t}}{\lambda}\right)+\left[1-\exp\left(\frac{-\alpha_{R}\beta_{R}\delta{t}}{\lambda}\right)\right]\frac{\bm{I}}{\beta_{R}}.\end{equation}
For the Giesekus fluid, we simply integrate~\eqref{eq:A_src} numerically, using
\begin{equation}\bm{A}^{n+1/2}=\bm{A}^{\star}-\frac{\delta{t}}{\lambda}f_{R}\left(\bm{A}^{\star}\right).\end{equation}
For low Reynolds number flows of Oldroyd B fluids, we found no significant difference in results obtained between analytic and numerical integration of~\eqref{eq:A_src}.
\item\emph{Step 8:} Finally, knowing $\bm{A}^{n+1/2}$, we store $tr\left(\bm{A}^{n+1/2}\right)$, and calculate the polymeric stress $\bm{\tau_{p}}^{n+1/2}$ using~\eqref{eq:tau_A}.
\end{itemize}

\subsection{Boundary conditions}\label{sec:bcs}

Second order accurate (in space) no slip wall boundary conditions are applied using mirror particles: for every particle $i$ within a distance $r_{s}$ of a wall boundary, a mirror particle $j$ is created such that $\bm{r}_{ij}$ is perpendicular to, and bisected by the boundary. Appropriate boundary conditions are then enforced by relating properties at particle $j$ to properties at particle $i$: for a no slip wall, $\bm{u}_{j}=-\bm{u}_{i}$, approximates $\bm{u}=\bm{0}$ on the boundary. For inflow and outflow boundaries we employ a moving ghost particle technique following~\cite{kunz_2016}. At inlets, we prescribe values of velocity and polymeric stress, and apply a Neumann condition for pressure. At outlets, a first order Neumann condition for velocity, pressure and polymeric stress is applied. At free surfaces, we require that
\begin{equation}\bm{\sigma}\cdot\bm{n}=0,\label{eq:fs1}\end{equation}
where $\bm{\sigma}=\bm{\tau}-p\bm{I}$ and $\bm{n}=\left[n_{x},n_{y}\right]^{T}$ is the unit surface normal (pointing out of the fluid). Normally in incompressible SPH,~\eqref{eq:fs1} is satisfied by explicitly setting $p=0$ on the free surface, and implicitly assuming that $\bm{\tau}=\bm{0}$ by construction of the gradient operators (see e.g.~\cite{fang_2006} for details). In two dimensions~\eqref{eq:fs1} may be expanded and rearranged to give:
\begin{subequations}
\begin{align}p&=2\eta_{s}\left[n_{x}^{2}\frac{\partial{u}}{\partial{x}}+n_{y}^{2}\frac{\partial{v}}{\partial{y}}+n_{x}n_{y}\left(\frac{\partial{u}}{\partial{y}}+\frac{\partial{v}}{\partial{x}}\right)\right]+n_{x}^{2}\tau_{p}^{xx}+2n_{x}n_{y}\tau_{p}^{xy}+n_{y}^{2}\tau_{p}^{yy}\label{eq:fs_norm}\\
  0&=2n_{x}n_{y}\left(\frac{\partial{u}}{\partial{x}}-\frac{\partial{v}}{\partial{y}}\right)+\left(n_{y}^{2}-n_{x}^{2}\right)\left(\frac{\partial{u}}{\partial{y}}+\frac{\partial{v}}{\partial{x}}\right)+\frac{1}{\eta_{s}}\left[n_{x}n_{y}\left(\tau_{p}^{xx}-\tau_{p}^{yy}\right)+\left(n_{y}^{2}-n_{x}^{2}\right)\tau_{p}^{xy}\right]\label{eq:fs_tang}\end{align}\end{subequations}
Whilst it is not possible in the SPH framework to explicitly satisfy both~\eqref{eq:fs_norm} and~\eqref{eq:fs_tang}, we improve upon the usual approach by ensuring the normal stress condition~\eqref{eq:fs_norm} is satisfied. This is achieved to first order in time by using the velocity gradients and polymeric stresses obtained at the end of the prediction step to calculate (using~\eqref{eq:fs_norm}) the required free-surface pressure, and specifying this pressure as an inhomogenous Dirichlet boundary condition in the PPE. 

\subsection{Free surface identification}\label{sec:fs}
In incompressible SPH it is necessary to identify free-surface particles in order to apply the boundary conditions in the PPE. Free-surface particles are identified as those satisfying the inequality
\begin{equation}\langle\nabla\cdot\bm{r}\rangle_{i}\leq1.4+0.5\max\left(\hat{C}_{i}-1,-0.2\right)\label{eq:fsd}\end{equation}
where $C_{i}$ is the particle concentration, given by
\begin{equation}C_{i}=\sum_{j\in\mathcal{N}_{i}}V_{j}W_{ij};\quad\hat{C}_{i}=N_{FP}\frac{C_{i}}{\displaystyle\sum_{k\in FP}C_{k}},\label{eq:conc}\end{equation}
in which $FP$ is the set of fluid particles, and $N_{FP}$ the number of fluid particles. The second term on the RHS of~\eqref{eq:fsd} relaxes the criterion in regions of high particle density, as might occur due to the compression of a surface near a stagnation point. This reduces free surface particle mis-labelling, although it does not prevent it entirely. Surface-normal vectors are calculated by taking the gradient of particle concentration:
\begin{equation}\bm{n}^{\star}_{i}=\frac{1}{h}\displaystyle\sum_{j\in\mathcal{N}_{i}}V_{j}\nabla{W}_{ij},\label{eq:surfnorm}\end{equation}
where the term $1/h$ normalises the magnitude of the vector relative to the resolution. Following~\cite{chow_2018}, the surface-normal vectors are then smoothed according to:
\begin{equation}\bm{n}_{i}=\displaystyle\sum_{j\in\mathcal{N}_{i}}V_{j}\bm{n}_{j}^{\star}\frac{W_{ij}}{C_{j}}\end{equation}
We denote the set of free surface particles as $FS$, and the set of near-surface particles, those which have one or more neighbours in $FS$, as $NS$.

\subsection{Shifting procedure}\label{sec:shift}

In Lagrangian SPH, particles follow streamlines, which leads to highly irregular and anisotropic particle distributions in flows with non-parallel streamlines. The deterioration of particle distributions reduces accuracy and leads to numerical instabilities. To remedy this, a stabilisation procedure is used, in which a quasi-regular particle distribution is maintained by the introduction of a shifting velocity. The shifting velocity is assigned to $\bm{u_{ps}}$ in~\eqref{eq:mom} and is (on average) small compared with the maximum fluid velocity, although the method ceases to be perfectly Lagrangian. We use a modified form of Fickian shifting, based on~\citet{lind_2012}, as follows.
\begin{equation}\bm{u}_{\bm{s},i}=\begin{cases}\mathcal{D}_{i}\nabla{C}_{i}&\forall{i}\in{I}\\\mathcal{D}_{i}\left(\bm{m}_{i}\cdot\nabla{C}_{i}\right)\bm{m}_{i}&\forall{i}\in{FS}\end{cases}\label{eq:shifting},\end{equation}
where $\bm{m}$ is a unit vector tangent to the free surface ($\bm{m}_{i}\cdot\bm{n}_{i}=0$), and $I$ is the set of internal (non-free surface) particles. The shifting velocity coefficient $\mathcal{D}$ is set as
\begin{equation}\mathcal{D}_{i}=\begin{cases}\frac{h^{2}}{4\delta{t}}\\4h\left(\left\lvert\bm{u}_{i}\right\rvert+0.2\max_{i}\lvert\bm{u}\rvert\exp\left(\frac{-5\left\lvert\bm{u}_{i}\right\rvert}{\max_{i}\left(\lvert\bm{u}\rvert\right)}\right)\right).\end{cases}\label{eq:shiftcoef}\end{equation}
In this work, the first definition is used for internal flows, and the second for flows with free surfaces. The exponential term in the second definition yields non-zero shifting velocities even in the region of stagnation points, eliminating the need for artificial stress terms used in other SPH schemes (e.g.~\cite{rafiee_2007}).
The term $\nabla{C}$ in~\eqref{eq:shifting} originates as the gradient of the particle concentration, and in our method is given by
\begin{equation}\nabla{C}_{i}=\displaystyle\sum_{j\in\mathcal{N}_{i}}\left(1+\frac{1}{4}\left(\frac{W_{ij}}{W_{ii}}\right)^{4}\right)\nabla{W}_{ij}\delta{V}+\alpha_{Ak,i}\displaystyle\sum_{j\in\mathcal{N}_{i}}Q_{ij}\frac{\bm{r}_{ij}}{\lvert\bm{r}_{ij}\rvert}\label{eq:gradC}\end{equation}
with 
\begin{equation}Q_{ij}=Q\left(q\right)=\begin{cases}\frac{1}{32}&q<0.5\\q^{3}\left(1-q\right)^{3}&0.5\le{q}<1\\0&\text{otherwise},\end{cases}\end{equation}
in which $q=2\lvert\bm{r}_{ij}\rvert/\delta{r}$, and 
\begin{equation}\alpha_{Ak,i}=\begin{cases}\frac{64}{\pi{h}}&\forall{i}\in{I}\\0&\text{otherwise}.\end{cases}\end{equation}
The final term in~\eqref{eq:gradC} is based on the surface tension model of~\cite{akinci_2013}, and acts as a repulsive force between particles. For confined flows with no free surface, this term is omitted. For free surface flows, this term acts to prevent near surface particles migrating onto the free surface, ensuring an improved particle distribution near free surfaces. We note that this final term is only beneficial for low Reynolds number flows ($Re<5$), where viscous forces prevent the repulsive force from causing the free surface to rupture. 

\section{Numerical results}\label{numres}
\subsection{Poiseuille flow}
Our first test problem is planar Poiseuille flow. The computational domain is a unit square, with no-slip walls at the upper and lower boundaries, and periodic boundary conditions applied at the left and right boundaries. We define the Reynolds and Weissenberg numbers in terms of the channel width $H=1$ and the steady-state velocity magnitude at the channel centre-line $U$, giving $Re=\rho{U}H/\eta_{0}$ and $Wi=\lambda{U}/H$. The results presented here are made non-dimensional with the steady state centre-line velocity $U$, the time-scale $H/U$, and the viscous stress $\eta_{0}U/H$. Here we investigate the performance of the method for a range of $Wi$, $\beta$ and constitutive models. The flow is driven by a body force $\bm{g}=\left[8,0\right]^{T}$. Except where explicitly stated, results were obtained with a Lagrangian scheme.

\subsubsection{Oldroyd B: convergence}
\begin{figure}
  \includegraphics[width=0.49\textwidth]{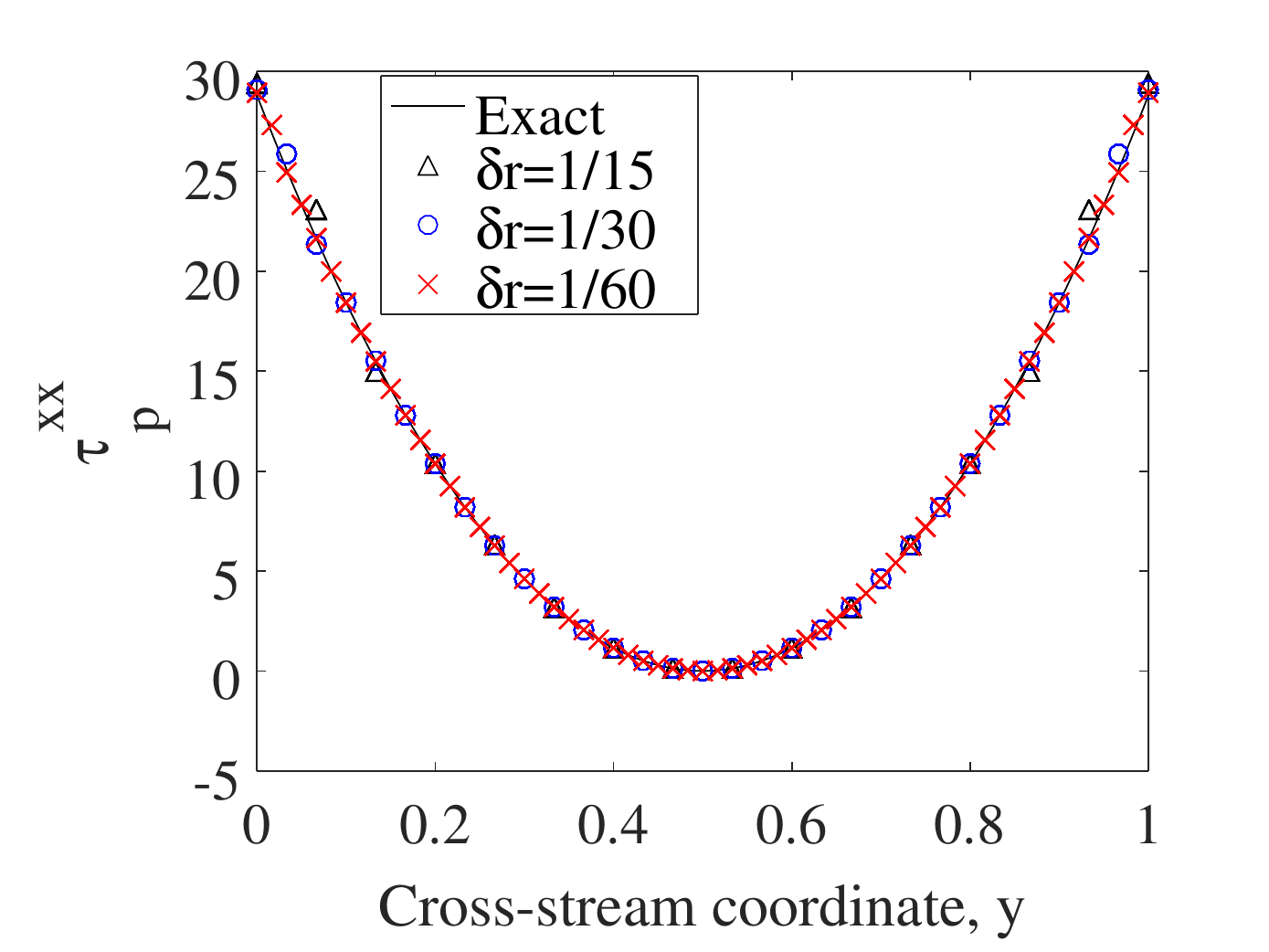}
  \includegraphics[width=0.49\textwidth]{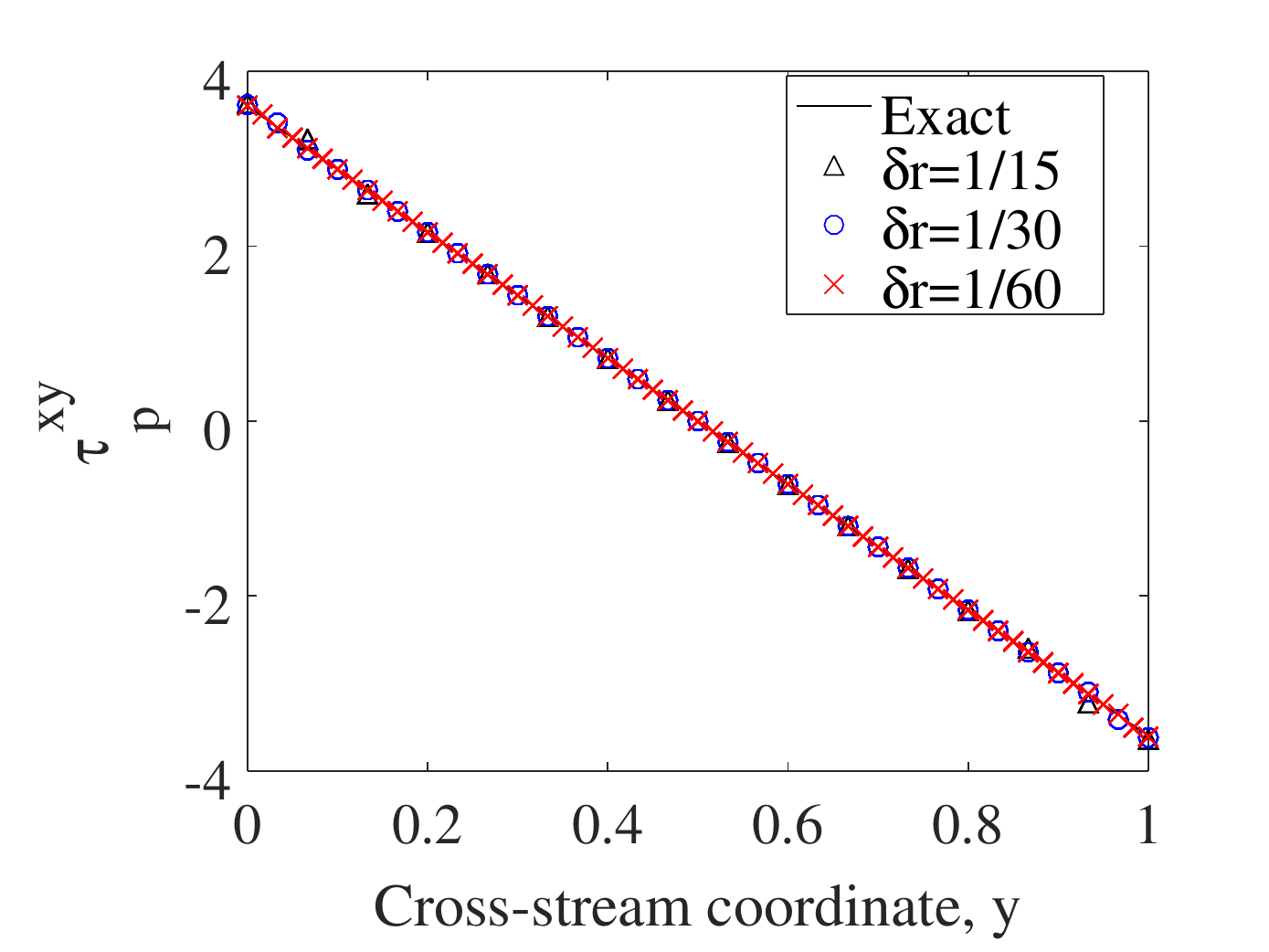} 
  \caption{Comparison of the analytical and numerical steady state stresses for the Oldroyd B fluid with $Re=1$, $Wi=1$ and $\beta=0.1$, for three resolutions. The left panel shows the normal stress $\tau_{p}^{xx}$, and the right panel shows the shear stress $\tau_{p}^{xy}$.\label{fig:ob_stresses}}
\end{figure}

\begin{figure}
\includegraphics[width=0.49\textwidth]{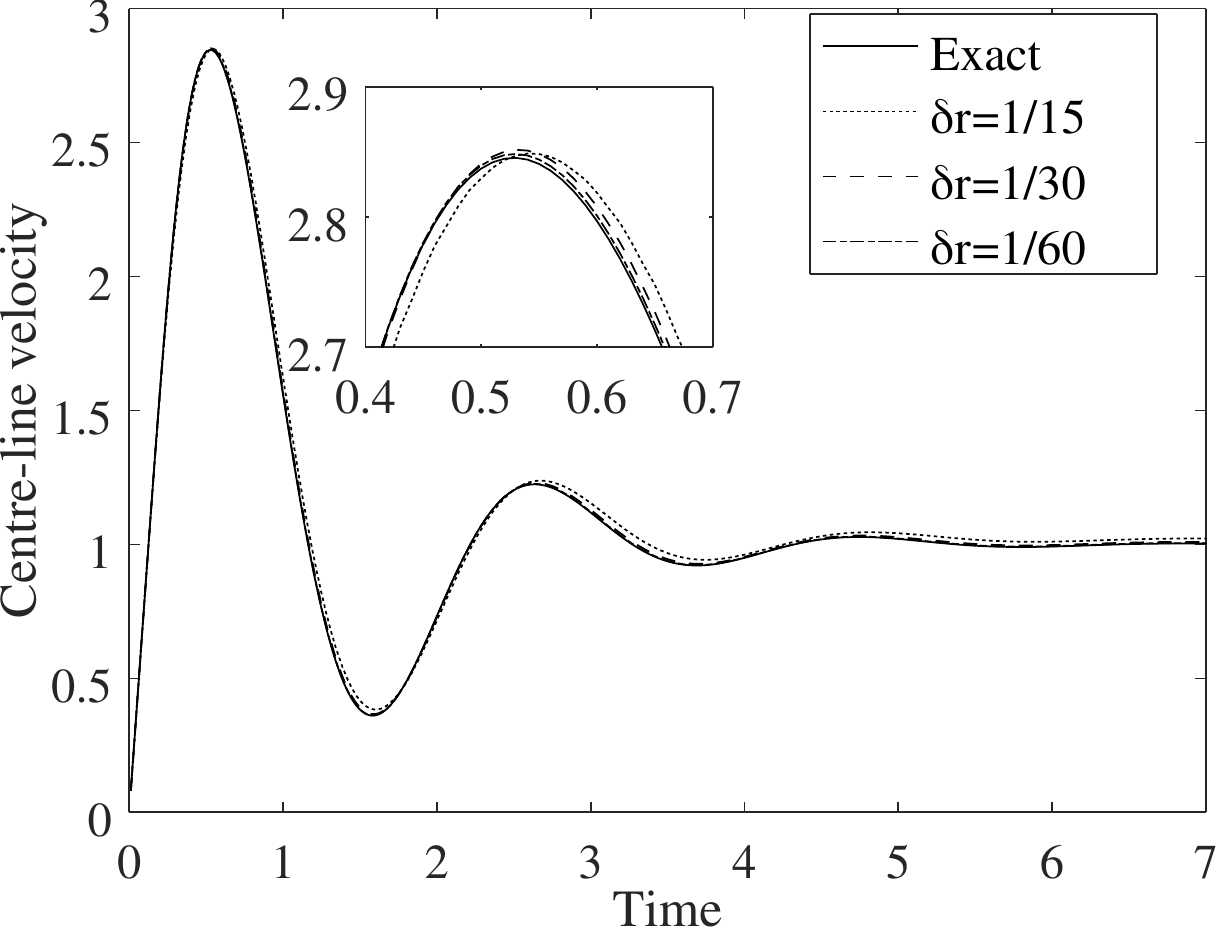}
\includegraphics[width=0.49\textwidth]{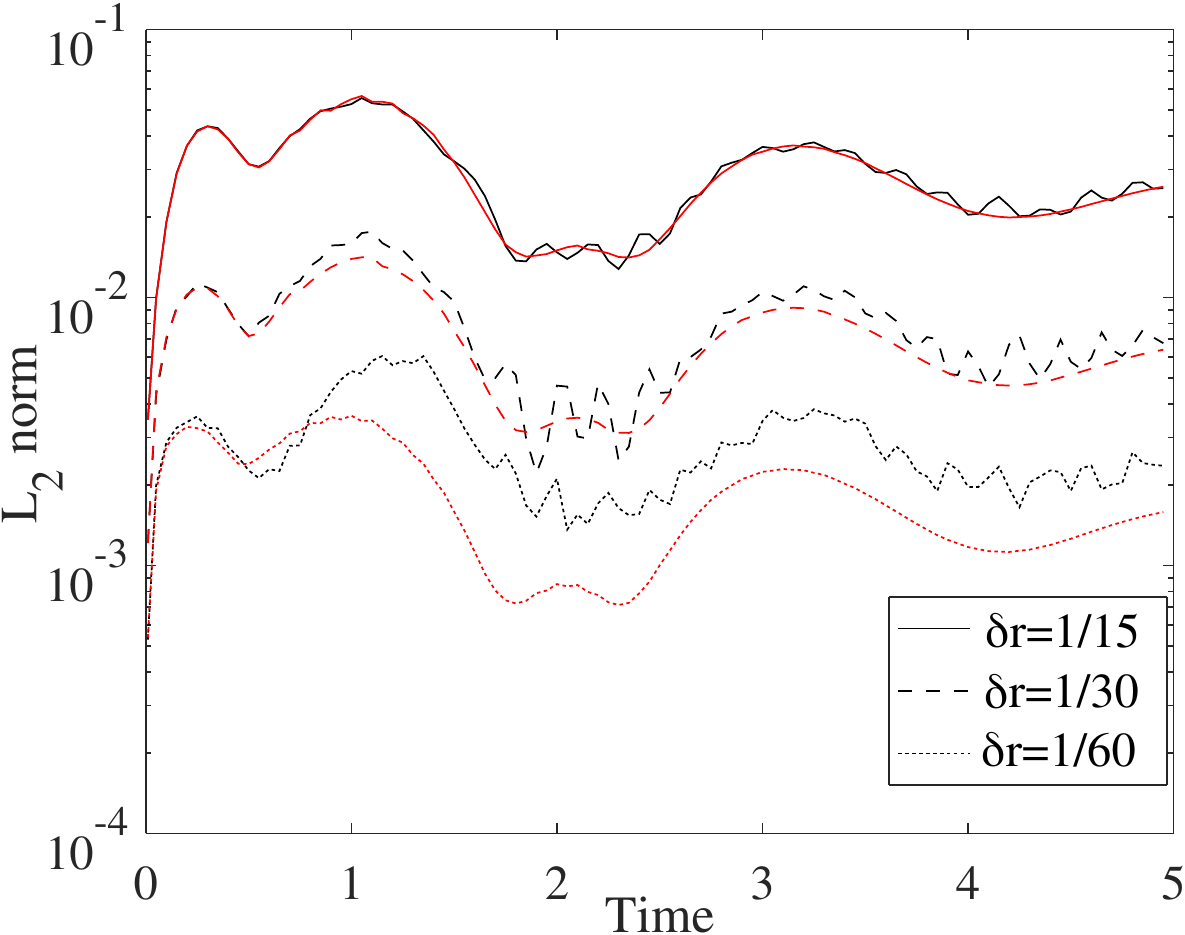}
\caption{Left panel: Variation of the centre-line velocity with time for the Oldroyd B fluid with $Re=1$, $Wi=1$, $\beta=0.1$, for three resolutions. The solid line shows the analytic solution. Right panel: Variation of the $L_{2}$-norm error of the velocity field with time for three resolutions, for Lagrangian (black lines) and Eulerian (red lines) SPH.\label{fig:ob_clv}}
\end{figure}

We first consider the convergence of the method for an Oldroyd B fluid at $Re=1$, $Wi=1$ and $\beta=0.1$. Figure~\ref{fig:ob_stresses} shows the steady state (taken at $t=15$) stress profiles for three resolutions, in comparison with the analytic solution of~\citet{waters_1970}. We see a good match even for a coarse resolution with only $15$ particles across the channel. Figure~\ref{fig:ob_clv} (left panel) shows the variation of the centre-line velocity with time for three resolutions. The numerical results (dashed lines) converge towards the analytic solution (solid line). 

To measure convergence we calculate the $L_{2}$-norm error of the streamwise component of the velocity field. Figure~\ref{fig:ob_clv} (right panel) shows the variation of this $L_{2}$-norm with time, for three resolutions, for both Eulerian (red lines) and Lagrangian (black lines) schemes. We see convergence, and for Eulerian SPH the convergence rate is $\mathcal{O}\left(\delta{r}^{2}\right)$. For Lagrangian SPH, there is increased temporal variation in the error, which is related to the changing particle distribution. Whilst at coarse resolution Lagrangian SPH converges at second order, as the resolution is increased, the order of convergence decreases (the dotted red line is below the dotted black line). In SPH spatial derivatives are approximated using a symmetric kernel function - the force on particle $i$ due to particle $j$ is equal and opposite to the force on particle $j$ due to particle $i$ - yielding second order convergence in the idealised case of a uniform particle distribution (as in the Eulerian case here). Whilst this approach makes SPH an attractively simple method to implement, for general particle distributions as the length scale of the resolution becomes small compared to the length scales of the flow features, the order of convergence decreases, and eventually the error may actually increase. Whilst up to second order is observed for a practical range of resolutions, in the present (Lagrangian) case, if the resolution is further doubled to $1/\delta{r}=120$, the errors are greater than for $1/\delta{r}=60$. This effect is well known in SPH, and the limiting error at fine resolutions is referred to as \emph{discretisation error}, which increases with particle disorder as analysed in detail in~\cite{quinlan,fatehi_2011}. As a consequence of this discretisation error, it is important in SPH simulations to choose a resolution which adequately resolves, but does not over-resolve, the flow features of interest.

\subsubsection{Oldroyd B: variation of $\beta$}

\begin{figure}
\includegraphics[width=0.49\textwidth]{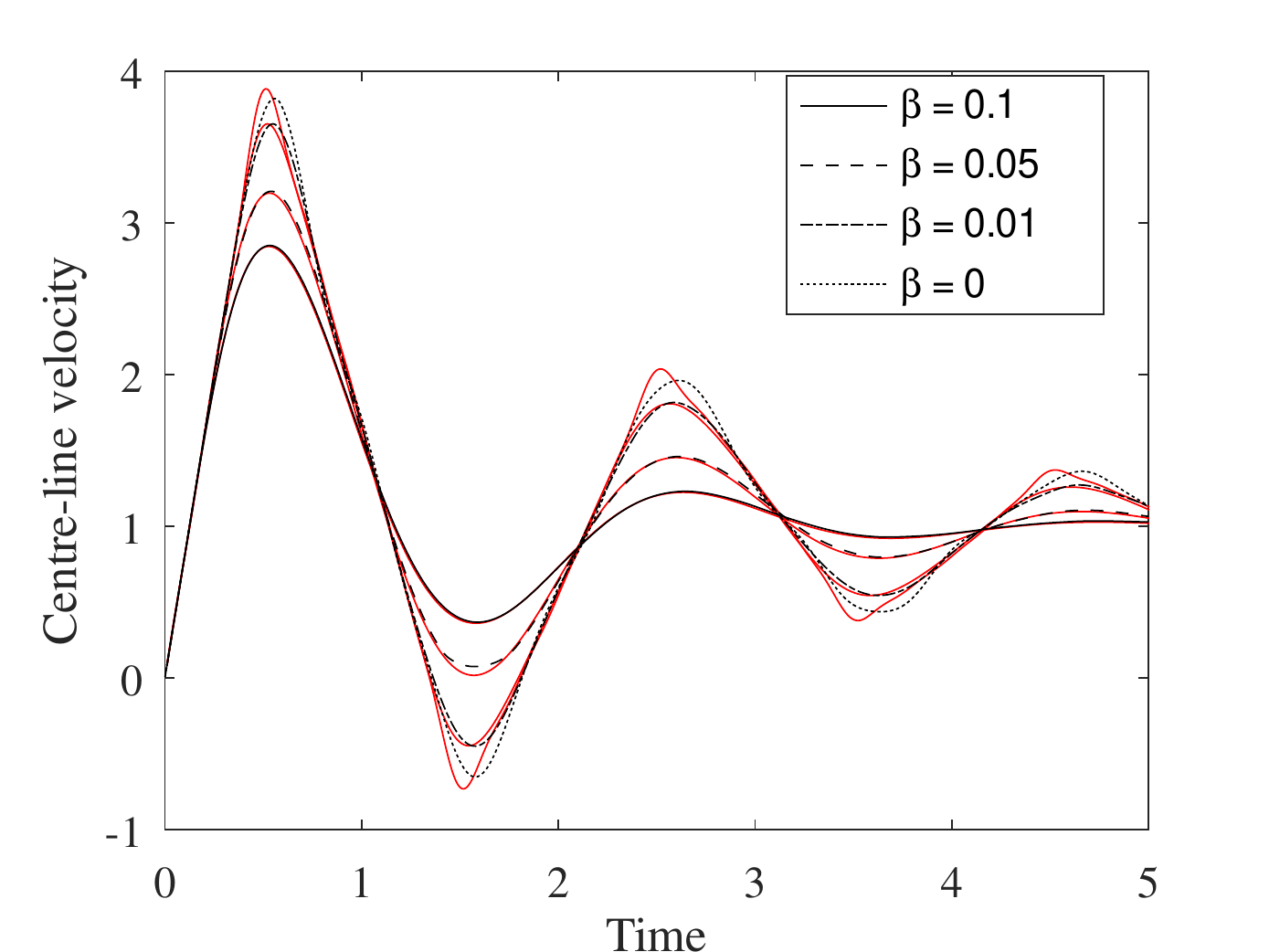}
\includegraphics[width=0.49\textwidth]{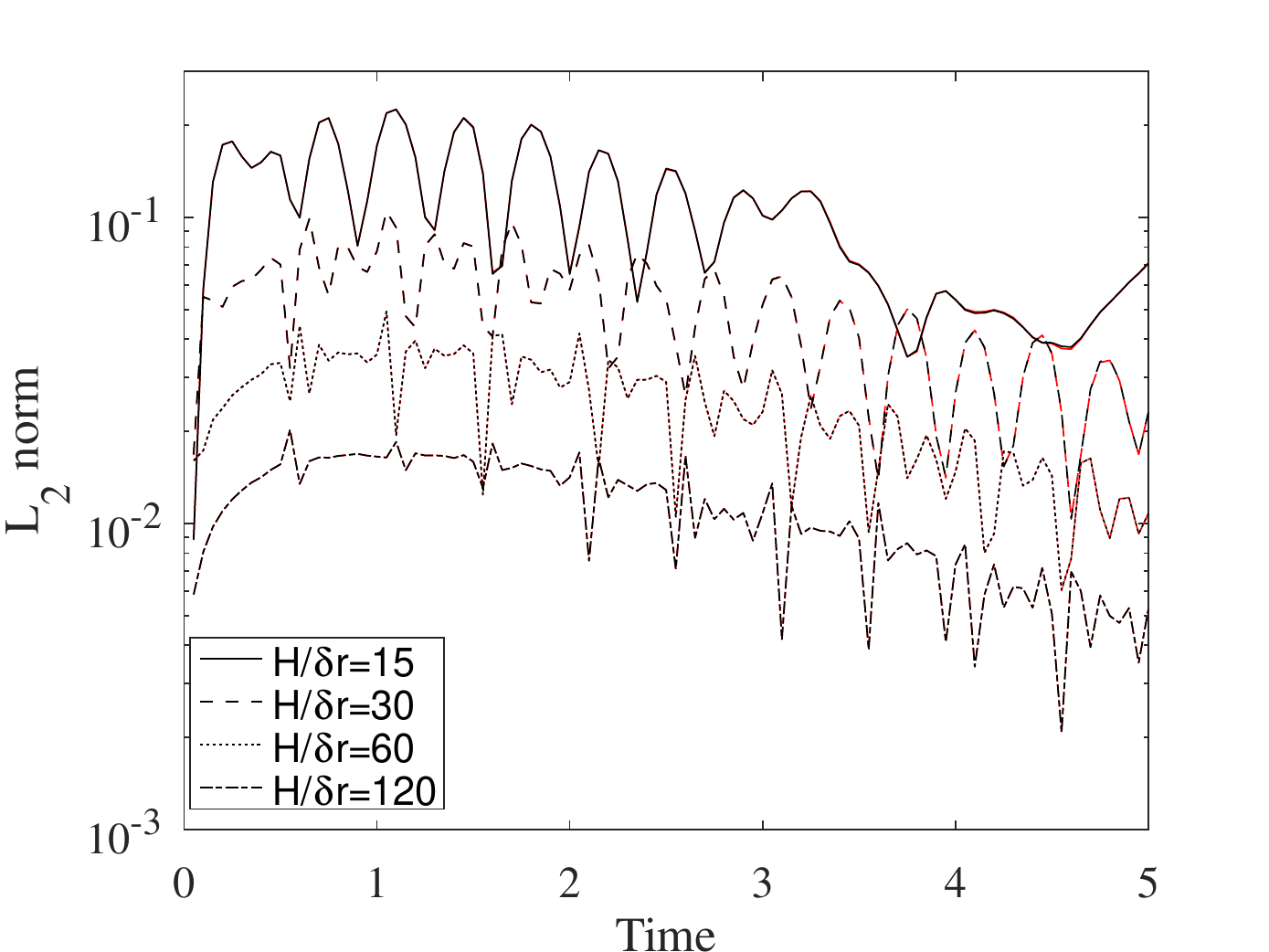}
\caption{left panel: Variation of the centre-line velocity with time for the Oldroyd B fluid with $Re=1$, $Wi=1$, with $\delta{r}=H/30$, for four values of $\beta$. The red lines show the analytic solution. Right panel: Variation of the $L_{2}$ norm error of the velocity field with time for $Re=1$, $Wi=1$ and $\beta=0$, for several resolutions, for Lagrangian (black lines) and Eulerian (red lines) simulations.\label{fig:ob_varb}}
\end{figure}

The left panel of Figure~\ref{fig:ob_varb} shows the variation of the centre-line velocity with $Re=1$, $Wi=1$ and for four values of $\beta$. For these cases we include the EVSS term, with $\alpha_{V}=0.01$. All cases were run with a resolution of $30$ particles across the channel. We see that as $\beta$ is reduced (the solvent viscosity decreases), we retain a relatively good match with the analytic solution, although the accuracy does deteriorate. For $\beta=0.01$ and $\beta=0.0$ there is a slight phase lag in the numerical simulations compared with the analytic solution. The deviation of the numerical results from the analytical solution increases as $\beta$ approaches zero. This is because the velocity profile contains higher modes, which are somewhat dissipated by the EVSS term in the simulations. Note that without the EVSS terms ($\alpha_{V}=0$) the simulation is unstable for $\beta=0$, and at late times deviates significantly from the analytic solution for $\beta=0.01$. We investigated varying $\alpha_{V}$, and found that for larger $\alpha_{V}$ the deviation from the analytical solution increases. The form of the viscosity added by the EVSS scheme is that in~\eqref{eq:lap}, whilst the cancelling term $-\nabla\cdot\left(\alpha_{V}\eta_{0}2\bm{D}\right)$ is based on two applications of~\eqref{eq:grad}. The two formulations differ in terms of the effective dissipation they introduce, with the latter increasing the effective stencil to include \emph{neighbours of neighbours}. For this reason, the diffusion introduced by the first formulation is not entirely removed by the second, a result which becomes apparent when $\alpha_{V}\ge\beta$.

The right panel of Figure~\ref{fig:ob_varb} shows the variation of the $L_{2}$ norm error of the velocity field with time, for $Re=1$, $Wi=1$ and $\beta=0$. The black lines indicate Lagrangian simulations, and the red lines indicate Eulerian simulations. In both cases we see convergence, and when averaged over time, the convergence rate is approximately $1.03$. We see that the errors for Lagrangian and Eulerian are (visibly) almost indistinguishable, in contrast to the errors shown in the right panel of Figure~\ref{fig:ob_clv}, for simulations with $\beta=0.1$. It is well known~\cite{quinlan_2006} that the error in SPH can be separated into a smoothing and a discretisation error, and that the discretisation error only dominates when the resolution becomes fine \emph{relative to the structure of the flow}. With $\beta=0.1$, the discrepency between the Lagrangian and Eulerian results in the right panel of Figure~\ref{fig:ob_clv} are a result of the flow being over-resolved, and discretisation error dominating.  With $\beta=0$, the coupling between the polymeric stress and the velocity field changes, resulting in transverse waves in the stress field. This results in finer structure in the velocity field, and even for the finest resolution tested ($H/\delta{r}=120$), the dominant error is due to inadequate resolution of the stress field, and not due to the particle disorder. This effect is also visible in the temporal variation of errors in the right panel Figure~\ref{fig:ob_varb}, which have a higher dominant frequency than for $\beta=0.1$ (in Figure~\ref{fig:ob_varb}), which is linked to the propagation of transverse waves, and not to the oscillations of the mass flow rate.

\begin{figure}
\includegraphics[width=0.49\textwidth]{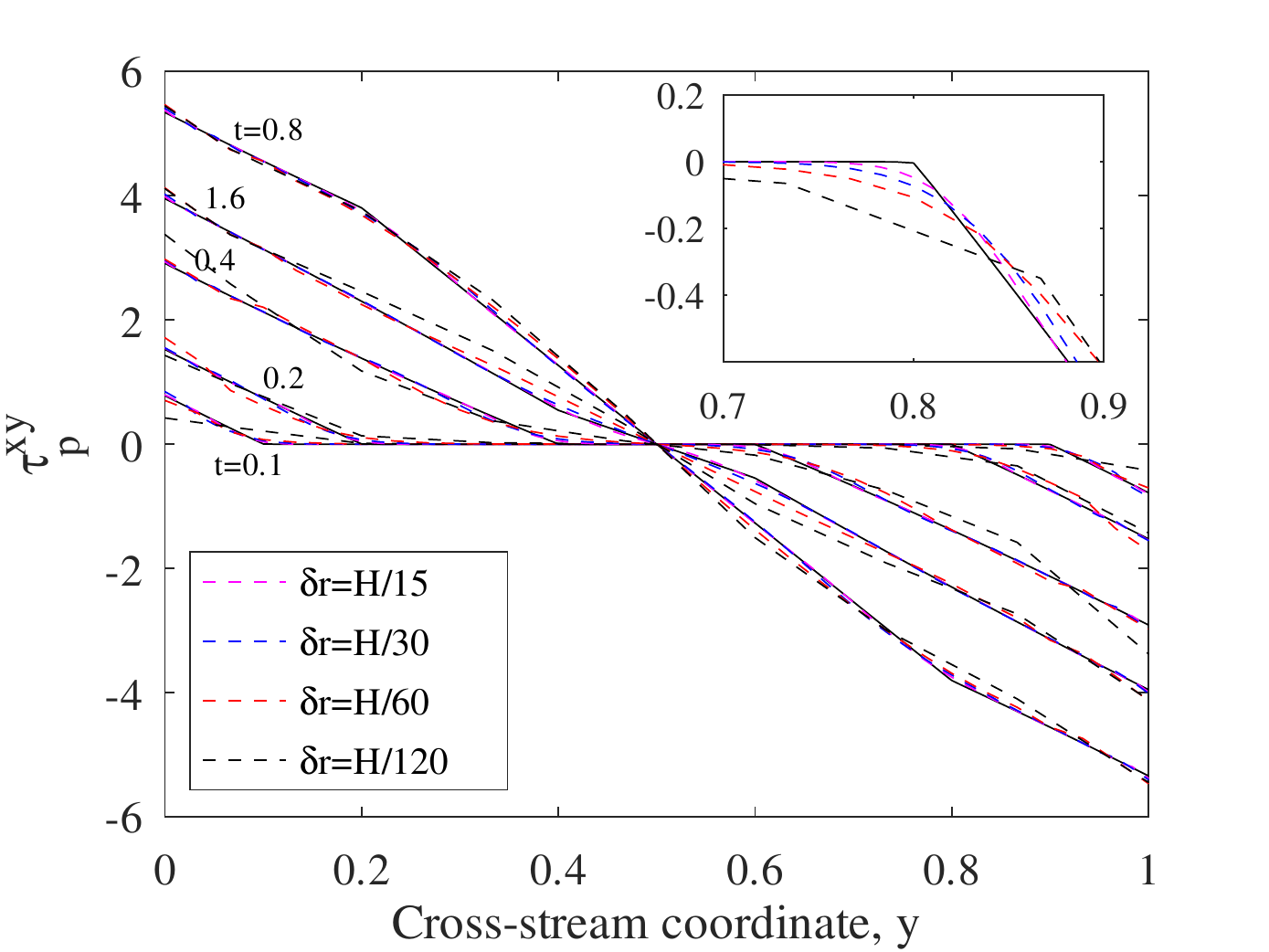}
\includegraphics[width=0.49\textwidth]{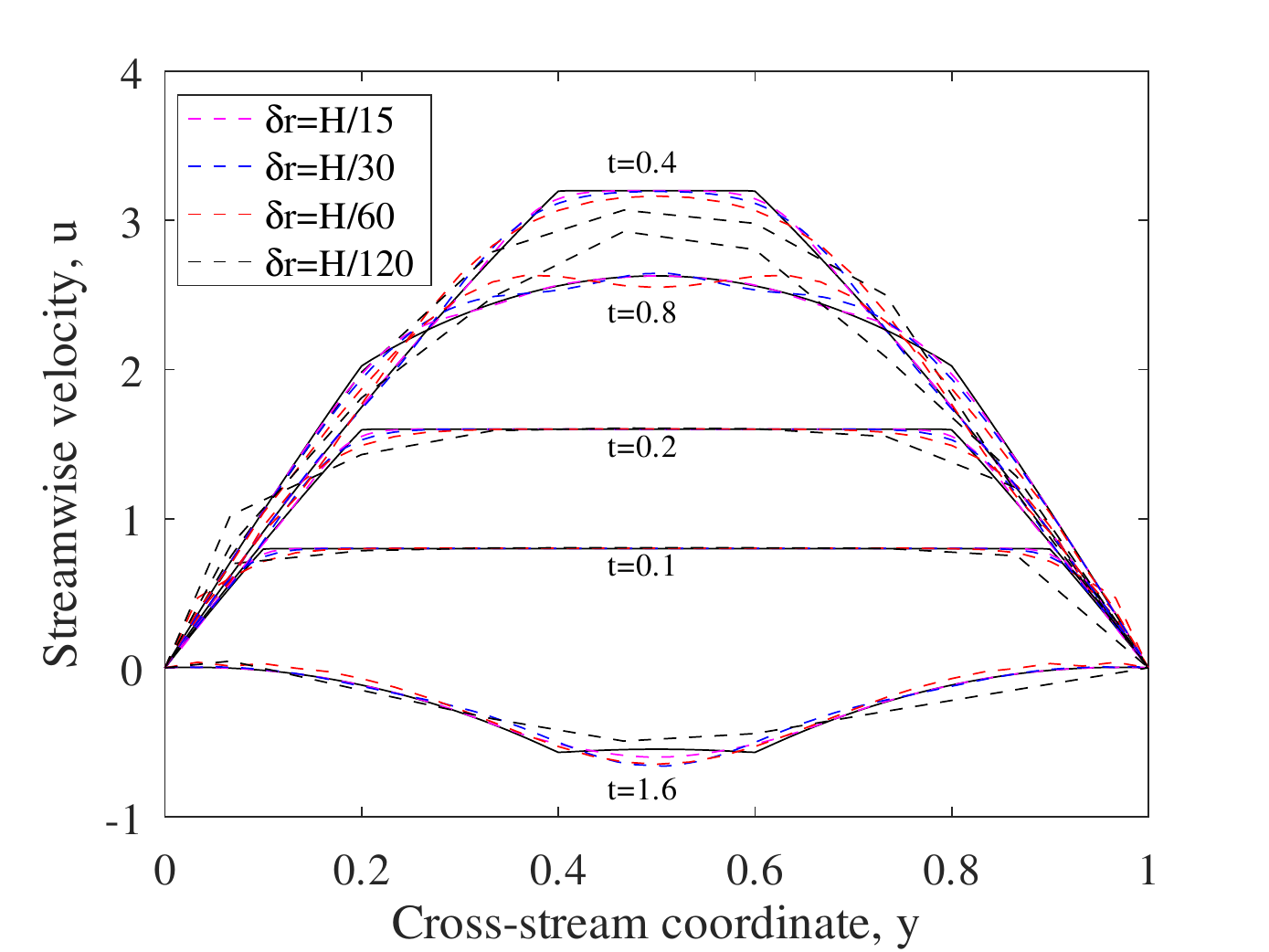}
\caption{Shear stress (left panel) and velocity (right panel) profiles for $Re=1$, $Wi=1$ and $\beta=0$, for a range of resolutions, at several times. The solid black lines indicate the analytical solution following~\cite{xue_2004}. The inset in the left figure shows the detail of the shear stress field at $t=0.2$.\label{fig:ucm}}
\end{figure}

Figure~\ref{fig:ucm} shows the shear component of the polymeric stress (left panel) and the streamwise velocity (right panel) for $Re=1$, $Wi=1$ and $\beta=0$, for several resolutions (different coloured dots) at several times during the transient start up of the flow. The solid black lines show the analytical solution following~\cite{xue_2004}. The transverse wave in the stress is clearly visible, though we see that in the numerical results the discontinuity in the stress gradient is smoothed. The transverse stress wave introduces a discontinuity in the velocity gradient, which can be seen in the right panel of Figure~\ref{fig:ucm}. Again, we see that the velocity field is smoothed relative to the analytic solution. In both the velocity and stress fields, the convergence of the numerical results towards the analytic solution is apparent, and the largest discrepancies occur around the transverse wave front. This further supports the observation based on Figure~\ref{fig:ob_varb} that the dominant error relates to the under-resolution of the wavefront, and the simulations here are not discretisation error limited.

\subsubsection{Oldroyd B: high $Wi$}\label{highwi}

\begin{figure}
\includegraphics[width=0.49\textwidth]{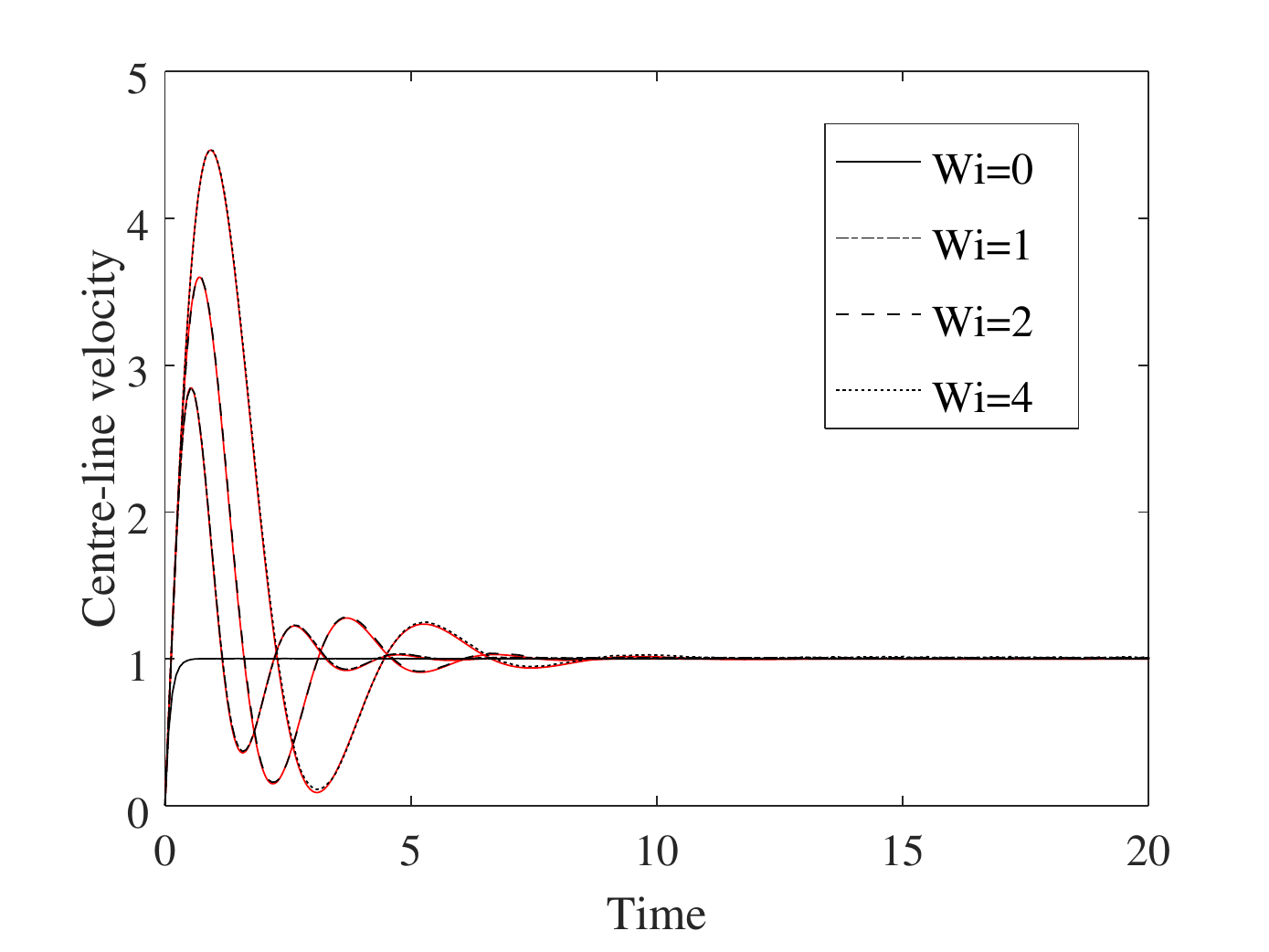}
\includegraphics[width=0.49\textwidth]{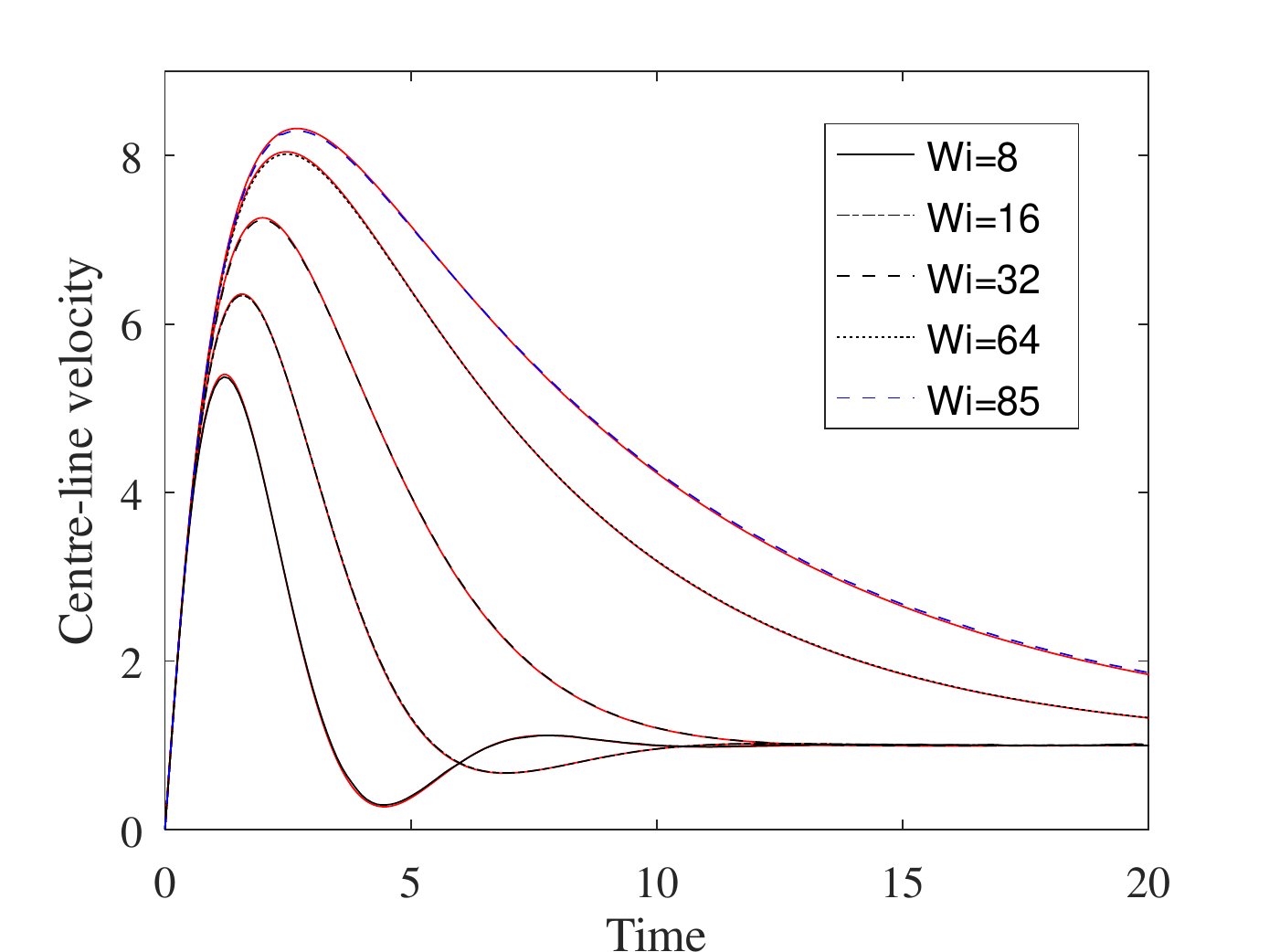}
\caption{Variation of the centre-line velocity with time for the Oldroyd B fluid with $Re=1$ and $\beta=0.1$, for increasing $Wi$. The red lines indicate the analytic solution, and the black lines indicate the numerical results.\label{fig:ob_clv_wi}}
\end{figure}

Next we hold $Re=1$ and $\beta=0.1$, and increase $Wi$. Figure~\ref{fig:ob_clv_wi} shows the variation of the centre-line velocity with time for increasing values of $Wi$, from $Wi=0$ to $Wi=85$. All results shown were obtained with a resolution of $30$ particles across the channel, using Lagrangian SPH. Up to $Wi\le{85}$, there is a good match with the analytical solution. \emph{The stability of the present Lagrangian scheme at $Wi=85$ represents a significant step change in the capabilities of SPH for modelling viscoelastic fluids, with previous works (e.g.~\cite{ren_2012}) being limited to $Wi\approx{1}$ and larger values of $\beta$ for internal shear flows.}  For higher $Wi$, the simulation breaks down, as a numerical instability arises, which we believe is due to the interaction between the polymeric stress and the particle distribution. Although Eulerian ISPH is more accurate at low $Wi$, we find it less stable as $Wi$ is increased, with the simulation crashing for $Wi\ge{16}$, and showing significant deviation from the analytical solution at late times for $Wi=8$. We postulate that the increased stability of Lagrangian over Eulerian ISPH is due to the absence of any special treatment of the advection term, a term which is included implicitly in the Lagrangian scheme through the particle motion. We observed that the eventual break down of the simulations is different for Lagrangian simulations - where an instability arises near the walls (discussed below) - and Eulerian simulations, where a chequerboard-type instability arises in the centre of the channel during the first peak in mass flow rate. 

Our scheme contains three elements designed to improve numerical stability: the log-conformation formulation, particle shifting, and the EVSS terms. We now investigate the effects of these different elements on numerical stability. We find that even in parallel viscoelastic flows, the particle shifting technique is essential to ensure stability; with $\bm{u_{ps}}=\bm{0}$ the simulations are unstable even for $Wi<0.1$, whilst with shifting, the maximum stable Weissenberg number is approximately $Wi=85$. This instability arises from the  at best first order consistency of the SPH operators, which result in non-uniform particle distributions to generating spurious streamwise forces (i.e. where $\partial\left(\cdot\right)/\partial{x}=0$ and $\partial\left(\cdot\right)/\partial{y}\ne0$, the SPH approximation of $\partial\left(\cdot\right)/\partial{x}$ may not be zero), in the direction of increasing particle disorder. For Poiseuille flow, with increasing $Wi$, the transverse stress gradients are steeper, and these spurious forces are larger. Particle shifting suppresses this instability, by introducing a force which promotes a uniform distribution. However, above $Wi=85$, these spurious forces overcome the stabilising effects of particle shifting and the simulation breaks down. If we replace the log-conformation formulation with a direct first-order Euler integration of~\eqref{eq:const_eq}, the maximum stable Weissenberg number reduces to $Wi=32$.

For the case of $\beta=0.1$, the EVSS scheme doesn't result in any increase of the maximum stable value of $Wi$, though, as shown above, it does enable smaller values of $\beta$ to be simulated. The maximum stable Weissenberg number decreases with reducing $\beta$, and for $\beta=0$, with $\alpha_{V}=0.01$, the simulation is stable and accurate at to $Wi=16$, but not at $Wi=32$.

\begin{figure}
\includegraphics[width=0.49\textwidth]{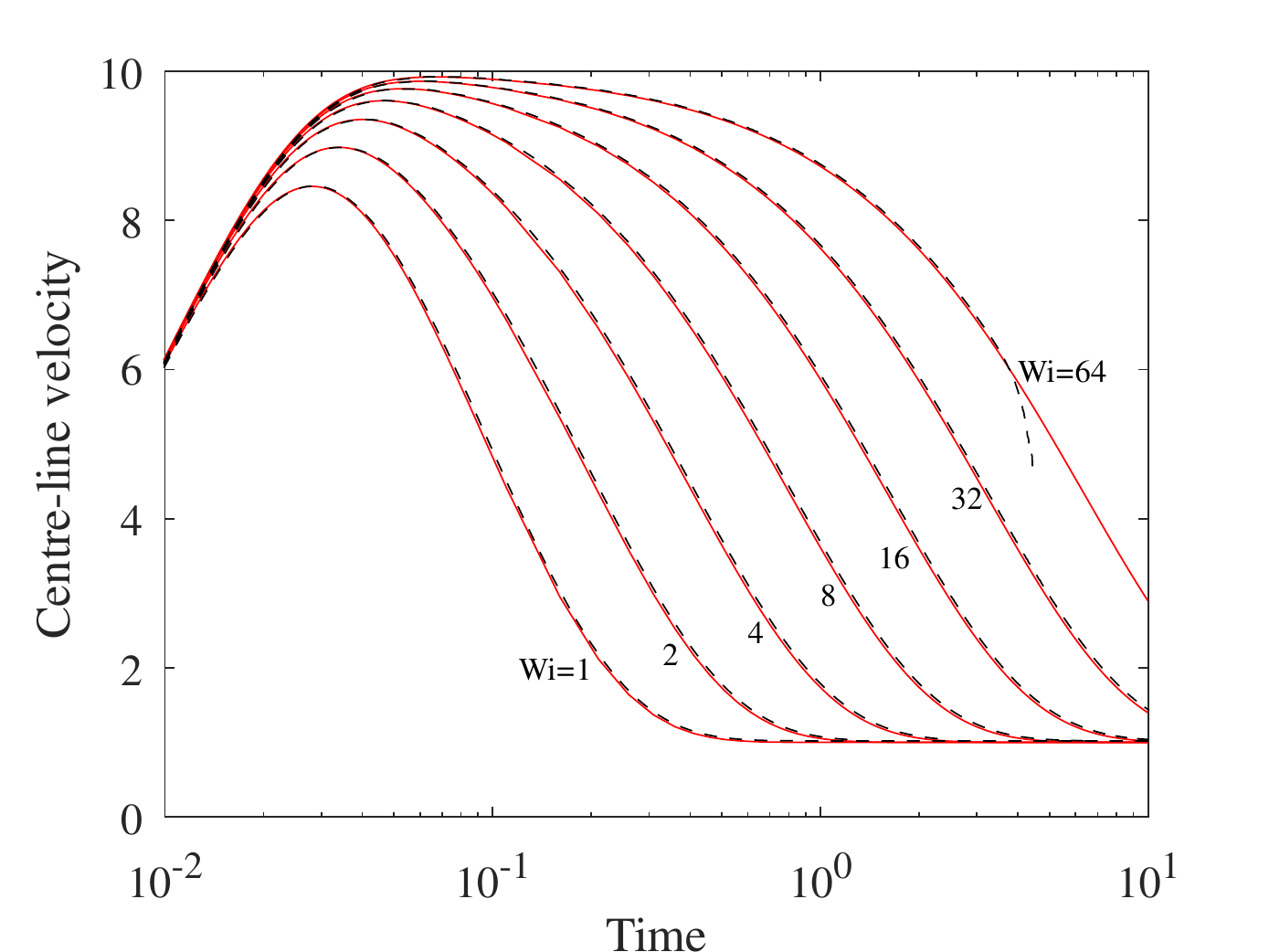}
\caption{Variation of the centre-line velocity with time for the Oldroyd B fluid with $Re=0.01$, $\beta=0.1$, for a range of Weissenberg numbers between $Wi=1$ and $Wi=64$. In all cases, $\delta{r}=H/15$. The red lines show the analytic solution, and the dashed black lines show the numerical results.\label{fig:creeping}}
\end{figure}

We next increase the elasticity number $El=Wi/Re$ by reducing the Reynolds number to $Re=0.01$, such that the stress gradient terms in~\eqref{eq:mom} become more dominant. Figure~\ref{fig:creeping} shows the time variation of the centreline velocity for a range of values of $Wi$, with $Re=0.01$ and $\beta=0.1$. Note the time is plotted on a log-scale for clarity, as the initial acceleration of the flow is extremely short lived compared with the subsequent decay. We find that the reduction in $Re$ (and corresponding increase in $El$) results in a lower maximum stable Weissenberg number. For values of $Wi$ up to $32$, there is a good match between the numerical results (dashed black lines) and the analytic solution (red lines). For $Wi=64$, the simulation breaks down during the later part of the decay (at approximately $t=4.5$). The maximum value of $Wi$ for which the simulation is stable in the long term was found to be approximately $Wi=40$. When the simulation does break down above this value of $Wi$, the instability again arises as a streamwise fluctuation near the wall, which cannot be suppressed by particle shifting. We note that in all the above cases using the log-conformation formulation, when the simulation breaks down, it is not due to the inability of our scheme to integrate the constitutive equation, but relates to the effects on particle motion of large stress gradients. In this regard, the implementation of the log-conformation formulation can be seen as highly beneficial here, removing one source of instability.

\subsubsection{Non-linear constitutive models}

\begin{figure}
\includegraphics[width=0.49\textwidth]{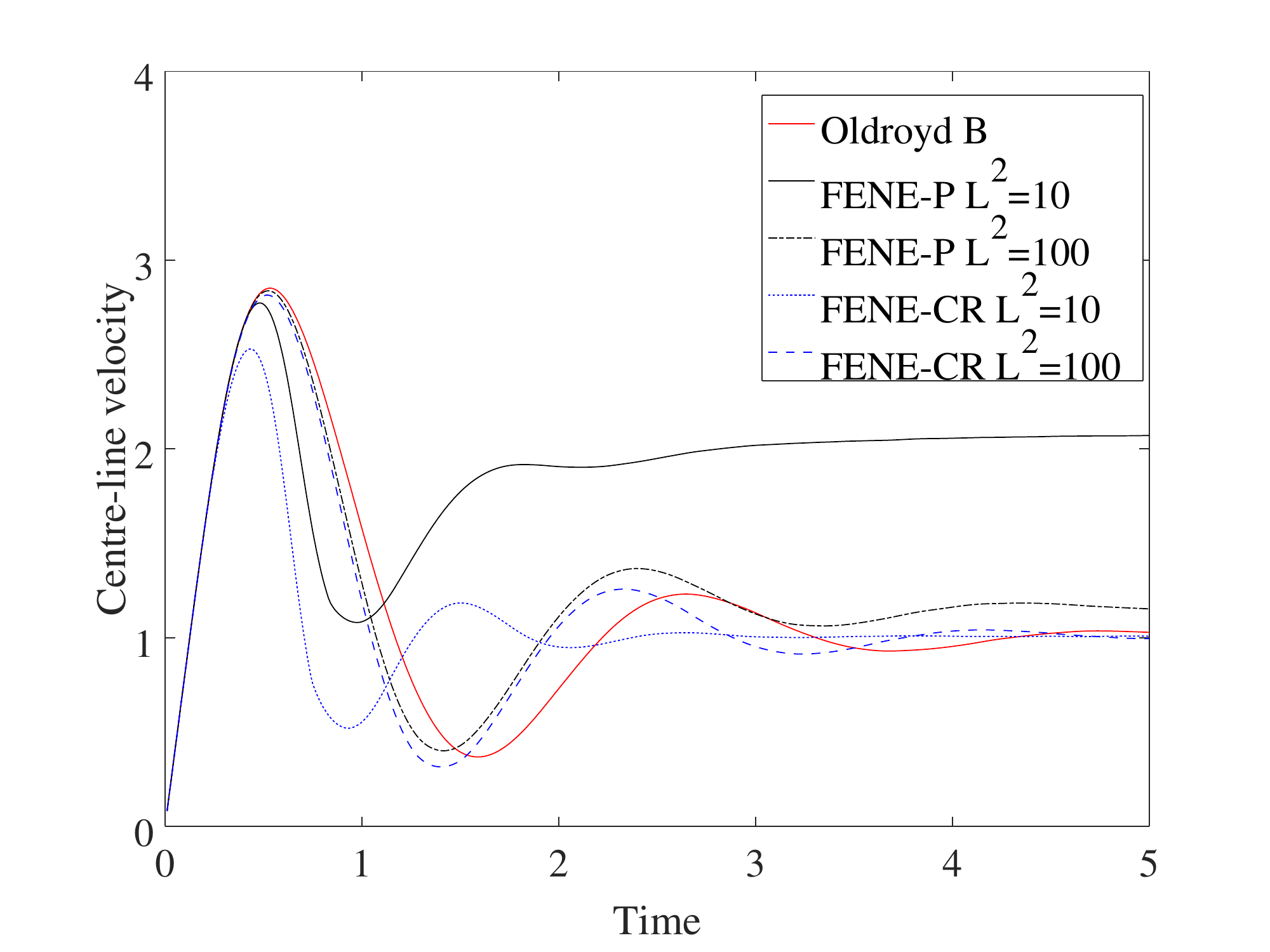}
\caption{Variation of the centre-line velocity with time for the FENE fluids with $Re=Wi=1$ and $\beta=0.1$. The black lines indicate the FENE-P model, the blue lines indicate the FENE-CR model, and the red line is the analytic solution for an Oldroyd B fluid.\label{fig:fene_clv}}
\end{figure}

Returning to the case with $Re=Wi=1$ and $\beta=0.1$, we now test our implementation of the FENE models. Figure~\ref{fig:fene_clv} shows the centre-line velocity obtained with both the FENE-P (black lines) and FENE-CR (blue lines) models, in comparison with the analytic solution for the Oldroyd B model (red line). We show results for two values of extensibility parameter $L^{2}=10$ and $L^{2}=100$. Firstly, we see that the steady state centre-line velocity for the FENE-CR model matches that of the Oldroyd B model, as expected. The FENE-P models show a degree of shear thinning, with larger steady state centre-line velocities for smaller $L^{2}$. For both models, the time to rebound is shorter than the Oldroyd B model, and shorter for smaller $L^{2}$. Setting $L^{2}=100$, we calculate the error in the stresses at $t=5$, and compare this with the analytic solution for the steady state stress given in~\cite{cruz_2005}. With $1/\delta{r}=30$, the relative $L_{2}$-norm errors are $5.2\%$ for $\bm{\tau_{p}}^{xy}$ and $3.4\%$ for $\bm{\tau_{p}}^{xx}$, with convergence rates of between $1$ and $2$. For smaller $L^{2}=10$ the numerical solution overestimates the stresses compared with the analytical solution, although this discrepency is expected. As discussed in~\cite{oliveira_2002}, the form of the FENE-P model used herein is only valid for $L^{2}\gg{2}$, and outside this range, an additional term ought to be included. For the PTT models we take the analytic solution of~\cite{oliveira_2002} for the steady state velocity profile. With $1/\delta{r}=30$, we find the relative $L_{2}$-norm is $0.8\%$ for the linear PTT model and $1.3\%$ for the exponential PTT model. Again we found approximately second order convergence with refinement of $\delta{r}$, up to the discretisation error limit at $1/\delta{r}=120$.

\begin{figure}
\includegraphics[width=0.6\textwidth]{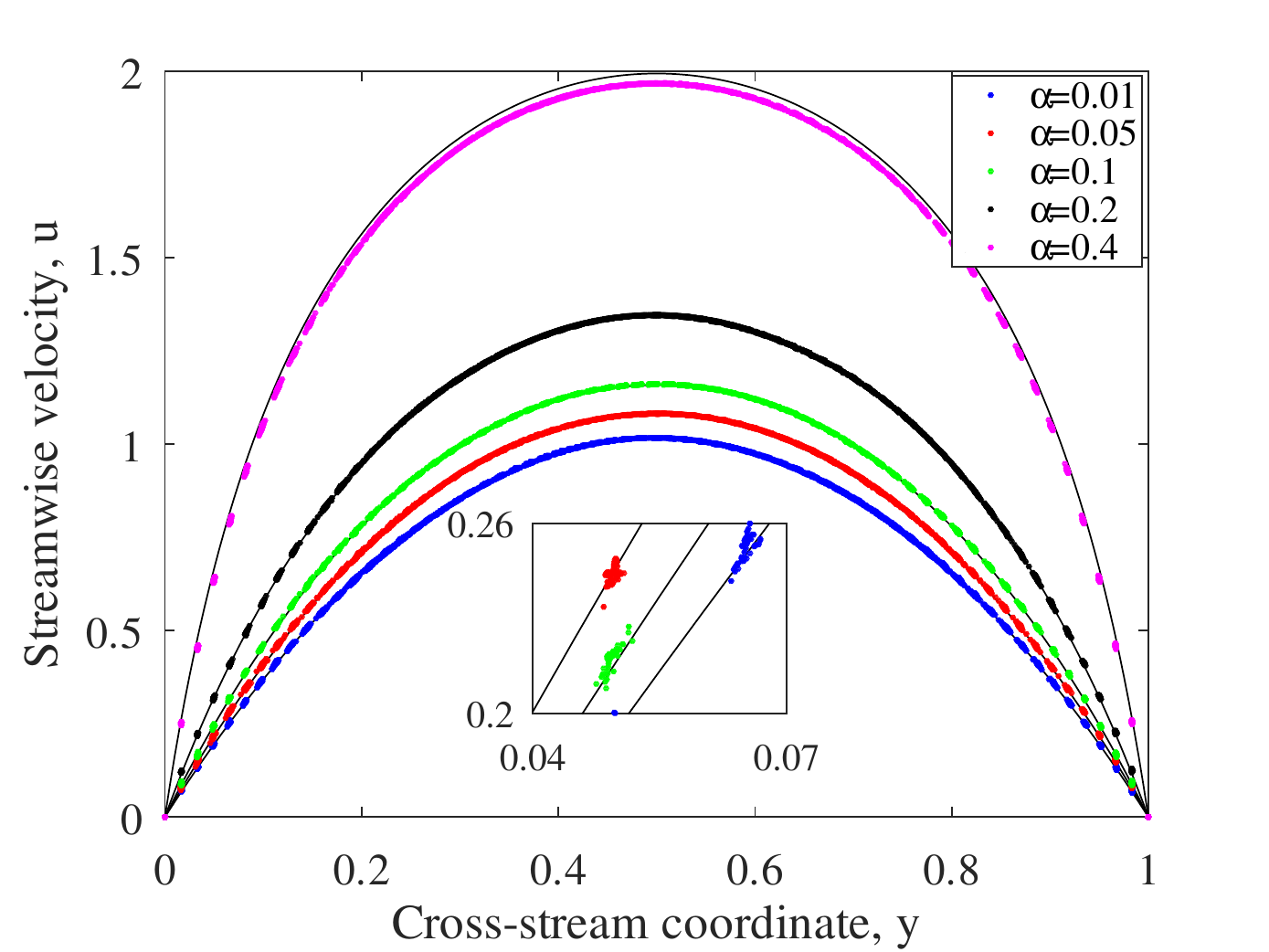}
\caption{Steady state velocity profiles for Poiseuille flow of a Giesekus fluid, with $Wi=0.25$, $Re=1$ and $\beta=0$, for a range of values of $\alpha$. Coloured dots denote numerical results, and black lines indicate the analytical solution. Note that results have been non-dimensionalised with steady state centre-line velocity for a UCM fluid. Each coloured dot represents an individual SPH particle, and each cluster of dots corresponds to $60$ particles along the channel length.\label{fig:giesekus_poiseuille}}
\end{figure}

For the Giesekus model, we consider the analytic solution of~\cite{yoo_1989} for a one-mode Giesekus fluid with zero solvent viscosity. Using a resolution of $60$ particles across the channel width, with $Re=1$, $Wi=0.25$, and $\alpha_{V}=0.1$, we take the steady-state solution as that at dimensionless time $20$, and have verified that the solution is steady with a run up to dimensionless time $140$. Figure~\ref{fig:giesekus_poiseuille} shows the steady state velocity profile for a range of $\alpha$. The coloured dots indicate the numerical results, whilst the solid lines indicate the analytic solution of~\cite{yoo_1989}. Results have been non-dimensionalised with the steady state centre-line velocity for a UCM fluid ($\alpha=0$). We observe a close match between the numerical results and the analytic solution, although this decreases as $\alpha$ approaches $0.5$. As $\alpha$ tends to $1/2$, the velocity gradient at the channel walls (in the analytical solution) tends to infinity. This asymptotic behaviour is not captured in our model due to viscosity introduced by the EVSS scheme (which is necessary to stabilise this case with $\beta=0$), and our numerical results underestimate the centreline velocity and channel wall velocity gradients. The inset of Figure~\ref{fig:giesekus_poiseuille} shows the spread of the numerical solution - each coloured dot represents an individual SPH particle, and each cluster of dots represents approximately $60$ particles along the length of the channel. Whilst the average of each cluster is very close to the analytic solution, there is some variation along the channel. This is due to the shifting procedure used to stabilised the ISPH algorithm, which combined with the Lagrangian form of the projection method introduces a small (first-order in space) error in the particle positions.

\subsection{Flow past a periodic array of cylinders\label{cyl}}

\begin{figure}
\includegraphics[width=0.99\textwidth]{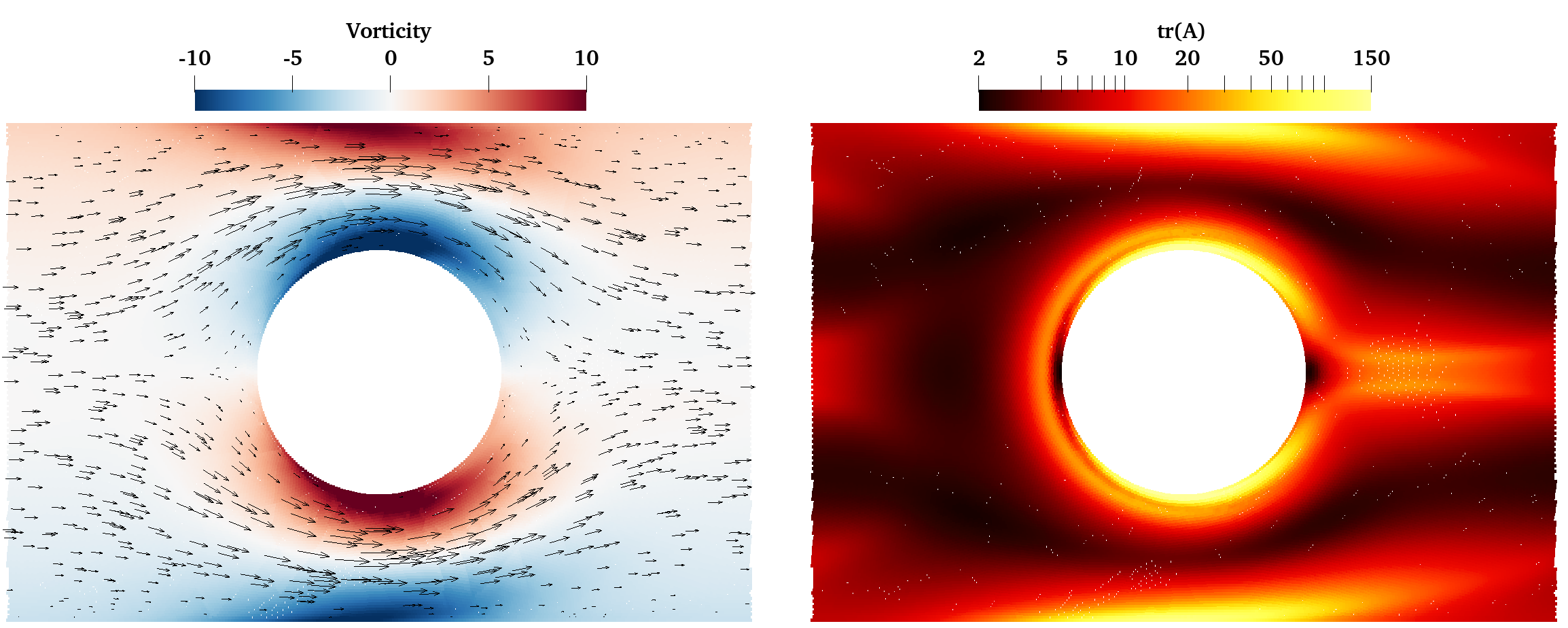}
\caption{Steady state vorticity field and velocity vectors (left) and conformation tensor trace (right) for flow past a periodic array of cylinders at $Wi=0.8$.\label{fig:cyl_fields}}
\end{figure}

We next consider the problem presented in~\cite{vazquez_2012}, of low $Re$ viscoelastic flow past an array of cylinders. As in~\cite{vazquez_2012}, the domain is rectangular, with length $6R$, height $4R$, and a central solid cylinder with radius $R=1$. At the upper and lower boundaries a no-slip wall condition is applied, whilst periodic boundary conditions are imposed in the horizontal direction. We use a resolution of $\delta{r}=6R/216$ which corresponds to resolution $P2$ in~\cite{vazquez_2012}. The fluid properties are $\rho=1$, $\eta_{0}=41.66$ and $\beta=0.59$. The flow is driven by a body force of $\bm{g}=\left(300,0\right)$, which results in an average velocity over the domain of $\left\langle{u}\right\rangle\approx1$, and a Reynolds number of $Re=2.4\times{10}^{-2}$. The Weissenberg number is $Wi=\left\langle{u}\right\rangle\lambda/R$, and by varying $\lambda$ we investigate different values of $Wi$. Whilst we found this case to be stable with $\alpha_{V}=0$, setting $\alpha_{V}=0.1$ reduced the disorder in the particle motions, yielding better distributions, without significantly affecting the resultant flow fields. 

\begin{figure}
\includegraphics[width=0.49\textwidth]{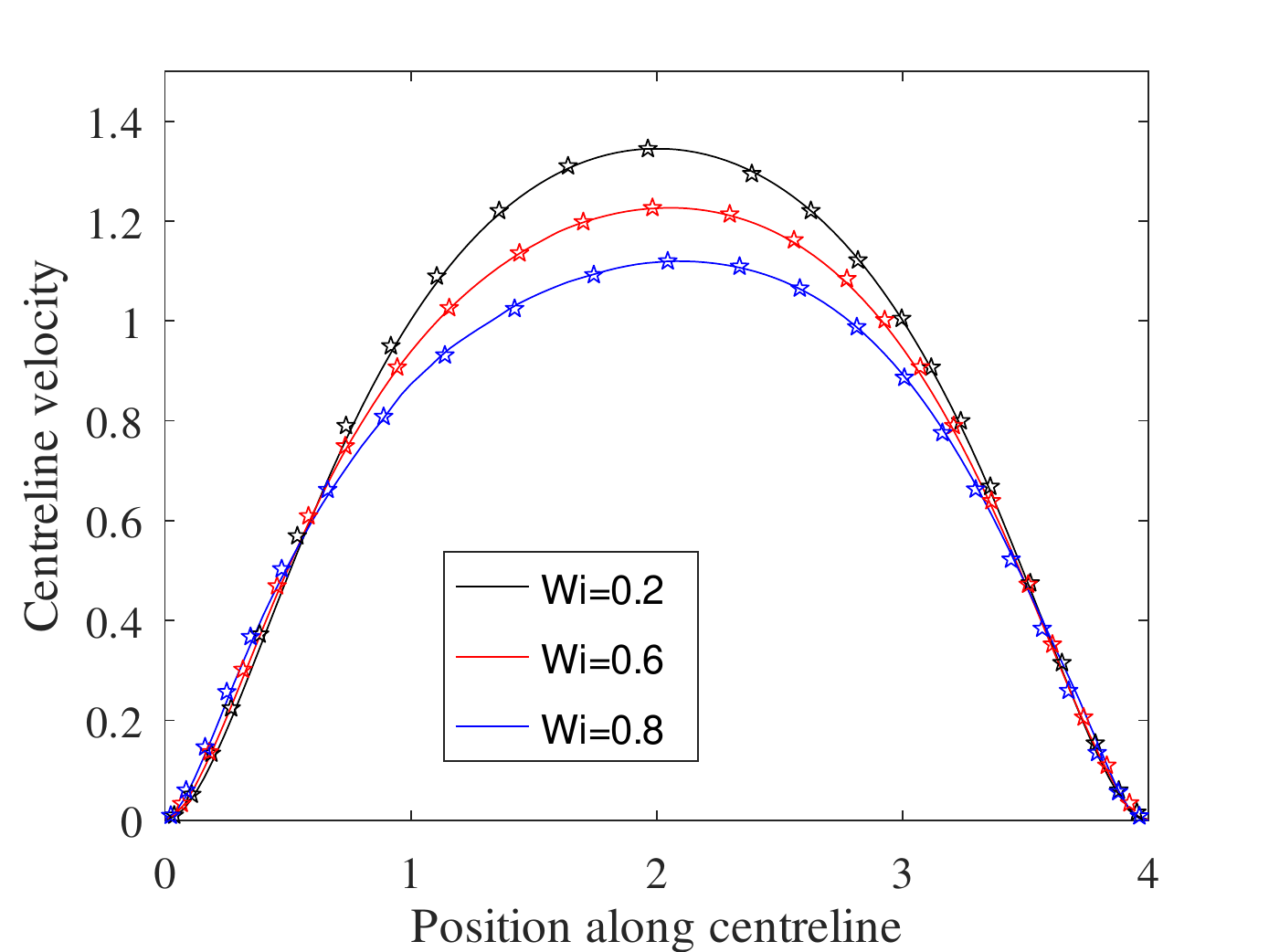}
\includegraphics[width=0.49\textwidth]{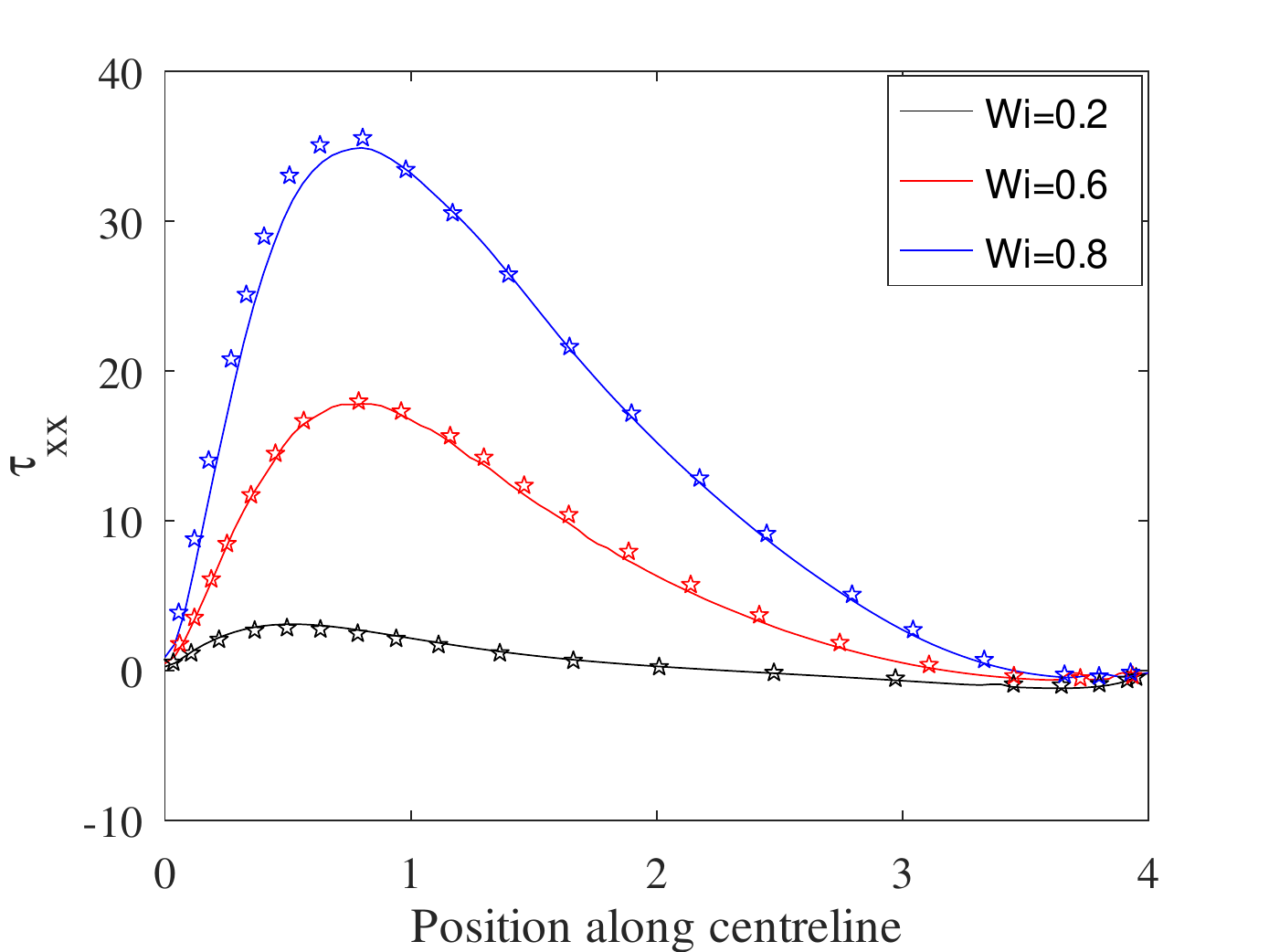}
\caption{Variation of the velocity (left) and stress component $\bm{\tau_{p}}^{xx}$ (right) profiles along the centreline of the channel between two cylinders, for a range of Weissenberg numbers. The solid lines indicate our ISPH results, whilst the symbols correspond to the results of~\cite{vazquez_2012}.\label{fig:cyl_centre}}
\end{figure}

Figure~\ref{fig:cyl_fields} shows the vorticity and conformation tensor trace fields for the steady solution with $Wi=0.8$. The assymetry (about a vertical line through the centre of the cylinder) is clear in both the vorticity and conformation tensor fields. A region of high internal molecular deformation is visible as a thin sheet wrapping round the cylinder and advected downstream along the channel centreline. Figure~\ref{fig:cyl_centre} shows the velocity (left) and stress component $\bm{\tau_{p}}^{xx}$ (right) profiles along the centreline of the channel. The coordinate $0$ corresponds to the right hand side of one cylinder, and the coordinate $4$ corresponds to the left hand side of the next cylinder. Solid lines indicate our ISPH results, whilst symbols indicate the results of~\cite{vazquez_2012}. The velocity profile has been non-dimensionalised with the average velocity $\left\langle{u}\right\rangle$, and the stress component $\bm{\tau_{p}}^{xx}$ has been non-dimensionalised by the solvent stress $\eta_{s}\left\langle{u}\right\rangle/R$. Results are shown for $Wi=0.2$ (black lines/symbols), $Wi=0.6$ (red lines/symbols) and $Wi=0.8$ (blue lines/symbols). Generally we see a good match with the results of~\cite{vazquez_2012}, with increasing asymmetry in the velocity profile, and increasing polymeric stresses, for increasing $Wi$. In~\cite{vazquez_2012} the flow was driven by a body force of unspecified strength, we infer set dynamically such that $\left\langle{u}\right\rangle\approx{1}$. In our method we apply a constant body force, with the value set (by numerical experiment) to yield $\left\langle{u}\right\rangle\approx{1}$ for $\lambda=0$. At non-zero $Wi$, the resulting $\left\langle{u}\right\rangle$ deviate from unity by a few percent, which we believe explains the discrepencies (e.g., the slightly lower peak value of $\bm{\tau_{p}}^{xx}$ in our results for $Wi=0.8$). It was noted in~\cite{vazquez_2012} that the determinant of the conformation tensor remained greater than $1$. We note here that the same applies in our scheme - $\det\bm{A}>{1}$ - and that in our scheme, this property is guaranteed by construction.

\begin{figure}
\includegraphics[width=0.49\textwidth]{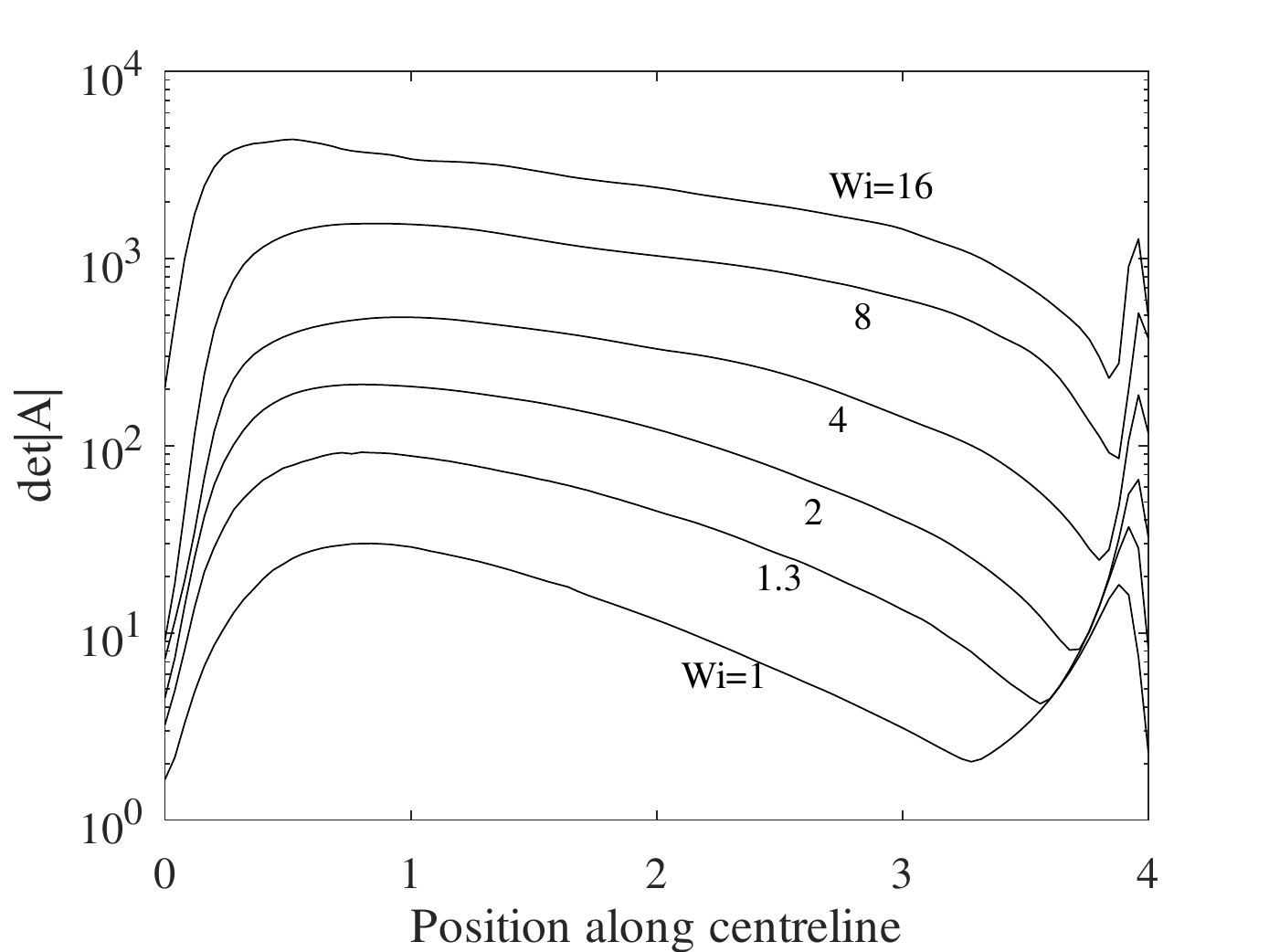}
\caption{Variation of the determinant of the conformation tensor $\det\bm{A}$ profile along the centreline of the channel between two cylinders, for a range of Weissenberg numbers. In all cases, the resolution is $\delta{r}=6R/216$.\label{fig:cyl_highwi}}
\end{figure}

As the Weissenberg number is increased, the layer of high internal molecular deformation just upstream of the leading edge of the cylinder (visible in the right panel of Figure~\ref{fig:cyl_fields}) becomes thinner, as does the region in the cylinder wake. At the same time, the magnitudes of the conformation tensor increase. As such, this is a challenging problem for the numerical method, as the stress gradients become extremely large. Figure~\ref{fig:cyl_highwi} shows the variation of the determinant of the conformation tensor $\det\bm{A}$ along the centreline of the channel, for increasing values of $Wi$ up to $Wi=16$. The simulations were run until the temporal variation of the mass flow rate was below $2\%$ over one dimensionless time unit, and the traces in Figure~\ref{fig:cyl_highwi} show the average values over a window of $\Delta{t}=3$ dimensionless time units thereafter. The first peak corresponds to the large streamwise polymeric stretching, and hence large values of $\tau_{p}^{xx}$ in the wake of the cylinder, whilst the narrow second peak corresponds to transverse deformation, and large values fo $\tau_{p}^{yy}$, just upstream of the cylinder. The increase in magnitude of $\det\bm{A}$ with increasing $Wi$ is visible in Figure~\ref{fig:cyl_highwi}, as is the reduction in the width of the peak just upstream of the cylinder. The simulation is stable for $Wi=16$, and we note that the maximum value of $Wi$ we are able to simulation is limited not by the stability of our scheme, but by the computational requirements of the problem. In this low Reynolds number regime, the value of the time-step is limited by the viscous constraint, and scales with $h^{2}$, whilst the cost per time-step scales with $h^{-2}\left(1-c\ln{h}\right)$, where $c$ is a constant and the logarithmic term originates from the increase in iterations required to solve the Poisson equation as the number of particles is increased. We see from Figure~\ref{fig:cyl_highwi} that the widths of the structures which we must resolve scale approximately with $1/Wi$, and furthermore, at higher $Wi$ the decay of transients is slower, and a steady state is reached later; the total physical time which must be simulated scales with $Wi$. Hence, the overall cost of conducting resolved simulations to a steady state scales with $Wi^{5}\left(1+c\ln{Wi}\right)$, and the simulations become prohibitively expensive for the present implementation, in the absence of spatially varying resolution. Nevertheless, we are able to conduct under-resolved simulations up $Wi=16$ with $\delta{r}=6R/216$. Although these simulations are globally stable and a statistically steady state is achieved, the under-resolution of the stress gradients gives rise to noise in the stress fields, the magnitude of which increases with increasing $Wi$. This can be seen in the trace for $Wi=16$ in Figure~\ref{fig:cyl_highwi} despite the averaging process, which is less smooth in the cylinder wake than at lower values of $Wi$.

\subsection{Drop impact}

\begin{figure}
\includegraphics[width=0.49\textwidth]{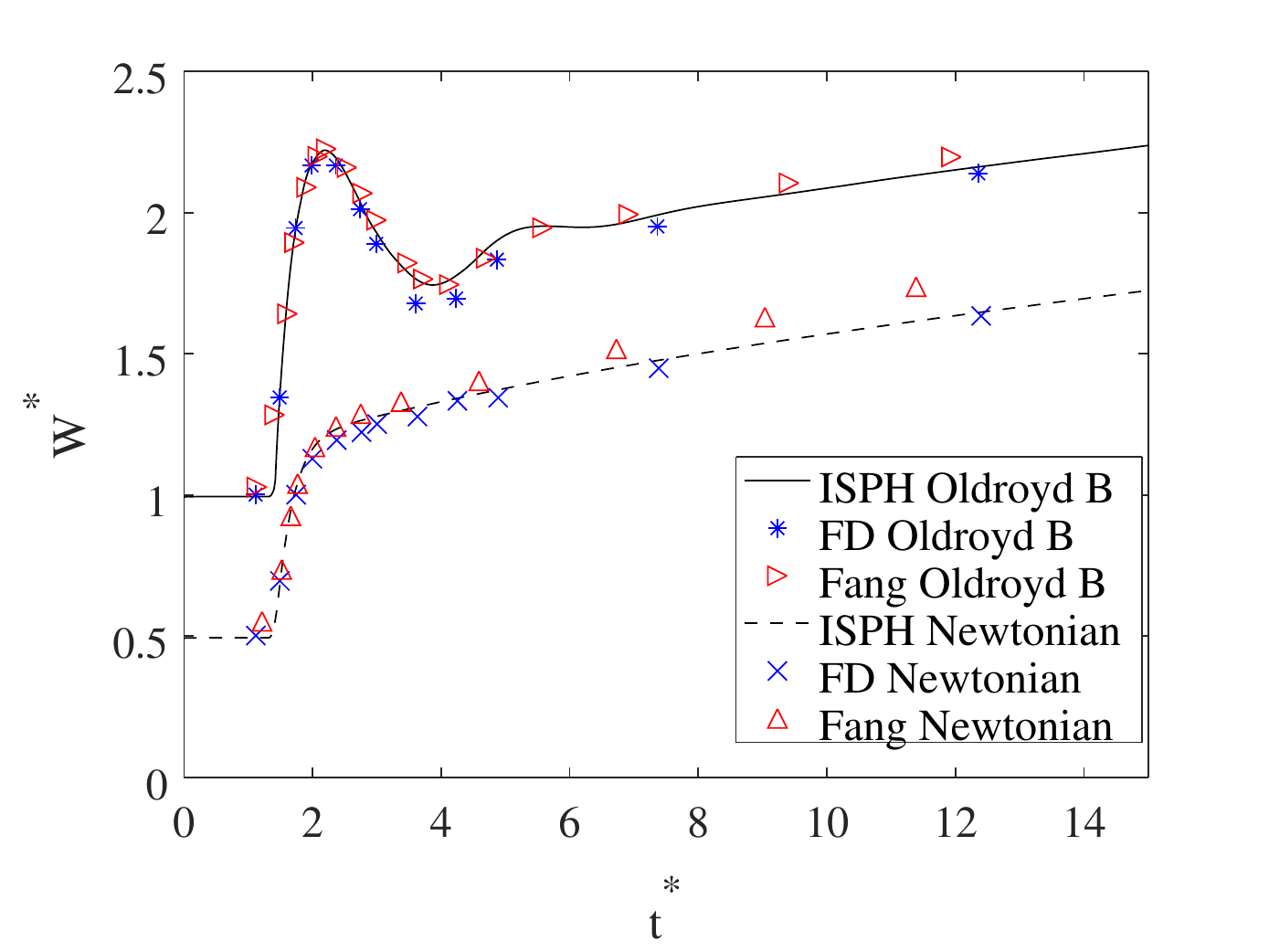}
\includegraphics[width=0.49\textwidth]{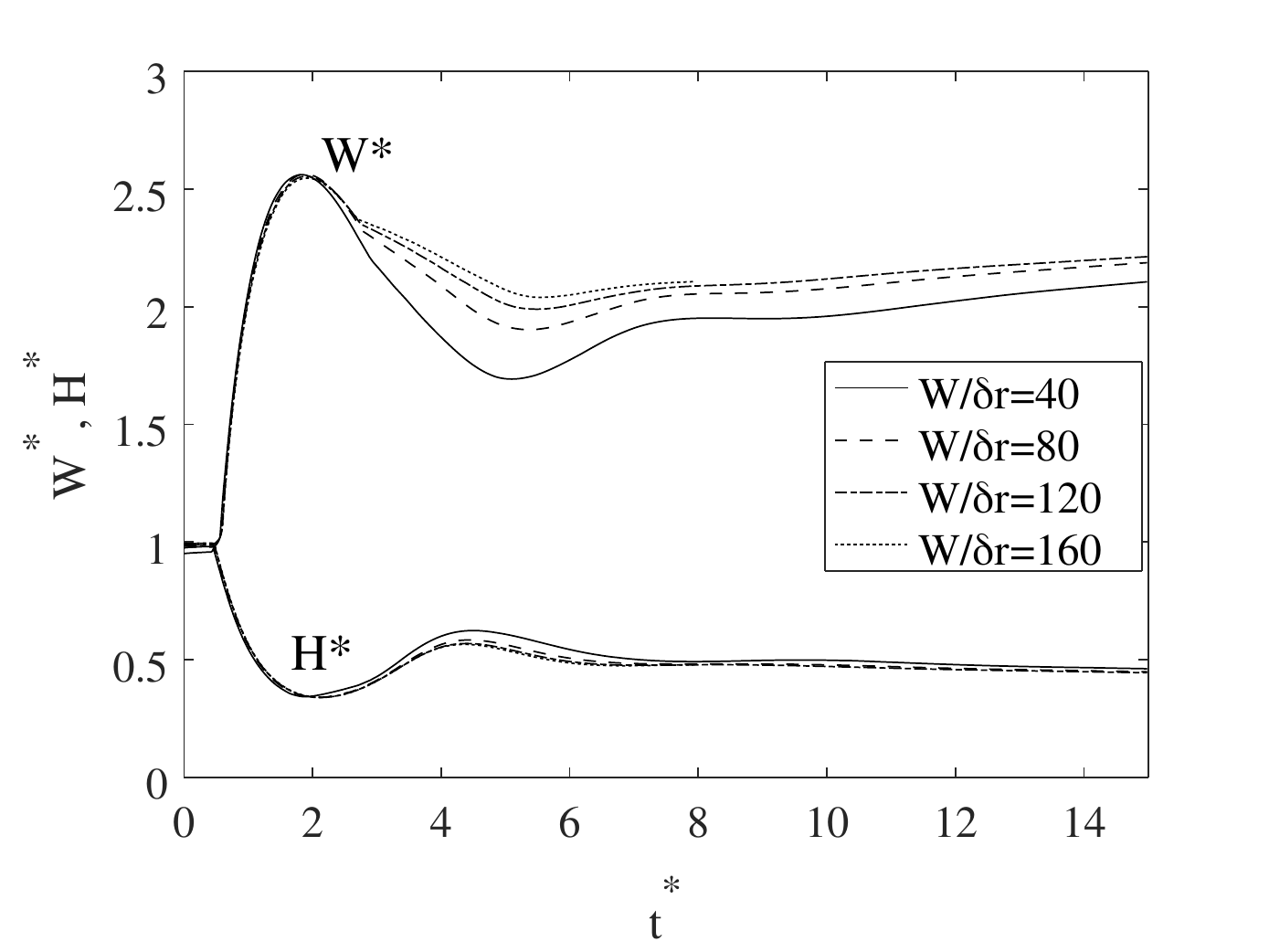}
\caption{Left panel: Variation of non-dimensional drop width $W^{\star}$ with non-dimensional time $t^{\star}$, for a Newtonian and an Oldroyd B drop. For the Olroyd B drop, the parameters are $De=1$ and $\beta=0.1$. Comparison is provided with the results of~\citet{tome_2002} (FD) and~\citet{fang_2006} (WCSPH). Note the results for the Newtonian drop have been shifted down by $0.5$ for clarity. Right panel: Variation of non-dimensional width $W^{\star}$ and height $H^{\star}$ of Oldroyd B drops with non-dimensional time $t^{\star}$, with increasing resolution.\label{fig:drops_ob_conv}}
\end{figure}

We consider the problem of a drop impacting on a solid surface. This problem has been studied by numerous authors, for finite difference~\cite{tome_2002}, and SPH~\cite{fang_2006,ren_2012,xu_2013} simulations. A circular drop of diameter $W=0.02$ is positioned with centre a distance $W$ above a flat solid plate. The origin is on the plate, directly below the centre of the drop. The drop initially has a velocity of $\left(u,v\right)=\left(0,-U_{d}\right)$, with $U_{d}=1$, and is subject to a gravitational acceleration of $g=9.81m/s^{2}$ downwards. The drop has density $\rho=1000kg/m^{3}$, and viscosity $\eta_{0}=5Pa\cdot{s}$. The polymeric stresses are initialised to zero. Using the characteristic length and time-scales $W$ and $W/U_{d}$, the Reynolds number is $Re=4$, and the Deborah number is $De=\lambda{U_{d}}/W$. In the following, we vary the viscosity ratio $\beta$, the relaxation time $\lambda$, the constitutive model, and the resolution $W/\delta{r}$. We calculate the non-dimensional drop width and drop height ($W^{\star}$ and $H^{\star}$) as the drop impacts on the plate and spreads. The non-dimensional time is $t^{\star}=tU_{d}/W$. Note that for these simulations we use the positive version of~\eqref{eq:grad} to evaluate the stress divergence, which although less accurate, and less stable at high $Wi$ for internal shear flows, is necessary in the presence of free surfaces.

\begin{figure}
\includegraphics[width=\textwidth]{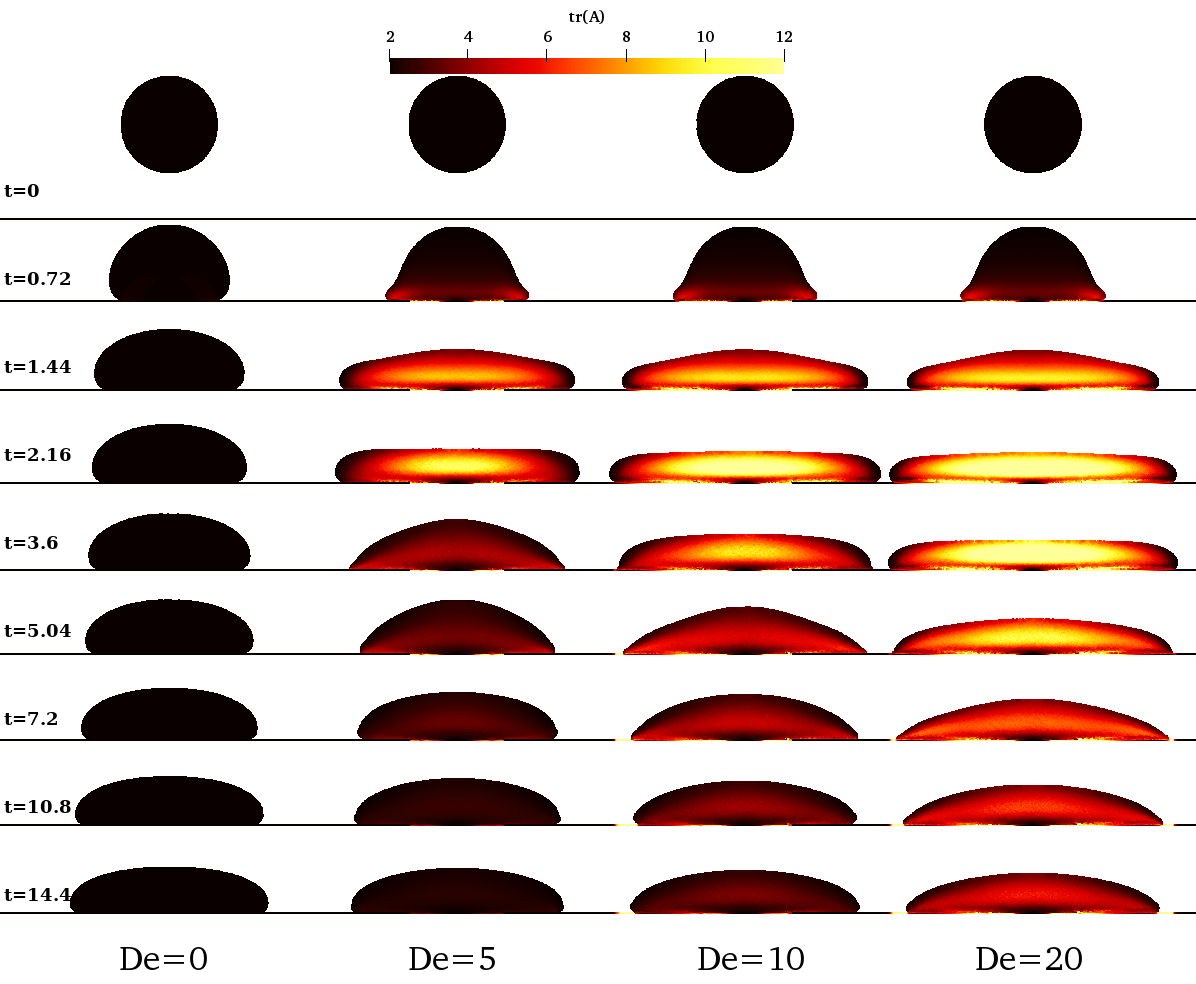}
\caption{Oldroyd B drops at a number of instants in time, for increasing values of $De$. Colour indicates the trace of the conformation tensor. Note that the colour scale is saturated at $t=2.16$ for $De=10$ and at $t=2.16$ and $t=3.6$ for $De=20$.\label{fig:drops_ob_varde}}
\end{figure}

The left panel of Figure~\ref{fig:drops_ob_conv} shows the variation of $W^{\star}$ with $t^{\star}$, and provides a comparison between our ISPH formulation, the results of the WCSPH scheme of~\cite{fang_2006}, and the finite difference (FD) scheme of~\cite{tome_2002}. For a Newtonian drop (lower line) our results match those of~\cite{tome_2002} well, whilst the WCSPH results of~\cite{fang_2006} overestimate the spreading at late times, although not significantly. For an Oldroyd B drop (with $De=1$ and $\beta=0.1$) (upper line), whilst the match is good generally, both SPH formulations underestimate the retraction in drop width (local minima at $t^{\star}\approx3.5$) compared to the FD simulations of~\cite{tome_2002}. This discrepancy is not due to the elastic stresses in the bulk of the drop, but is because the SPH formulations predict a smaller contact angle at drop/solid interface than the finite difference scheme of~\cite{tome_2002}. Note, none of the three methods being compared contain a model for surface tension. The right panel of Figure~\ref{fig:drops_ob_conv} shows the variation of $W^{\star}$ and $H^{\star}$ with non-dimensional time $t^{\star}$ as the resolution is refined, from $W/\delta{r}=40$ to $W/\delta{r}=160$. We see convergence with increasing resolution. There is a significant discrepency between the coarse and finer resolutions for $W^{\star}$ around $t^{\star}\approx5$, as the contact between the plate and the fluid and is under resolved for $W/\delta{r}=20$. Note that although surface tension is not included in the present model, the contact angle changes with time, and is dependent on the stress/conformation tensor. Including surface tension in the present ISPH formulation is an area of ongoing research for the authors.  We see convergence rates of approximately $2$ towards the finest resolution. In the following we use a resolution of $W/\delta{r}=120$ whilst varying $De$ and $\beta$.

\begin{figure}
\includegraphics[width=0.49\textwidth]{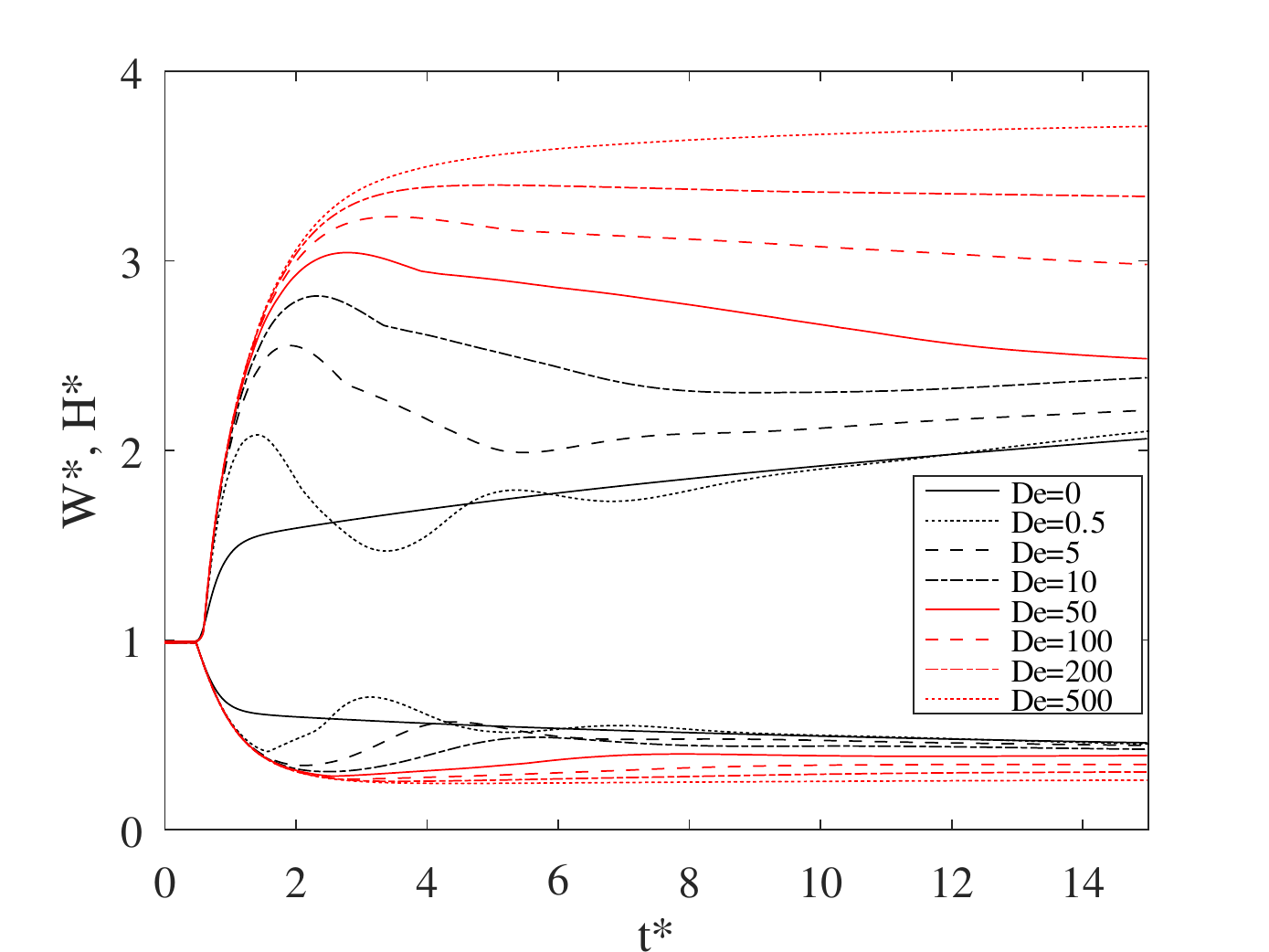}
\includegraphics[width=0.49\textwidth]{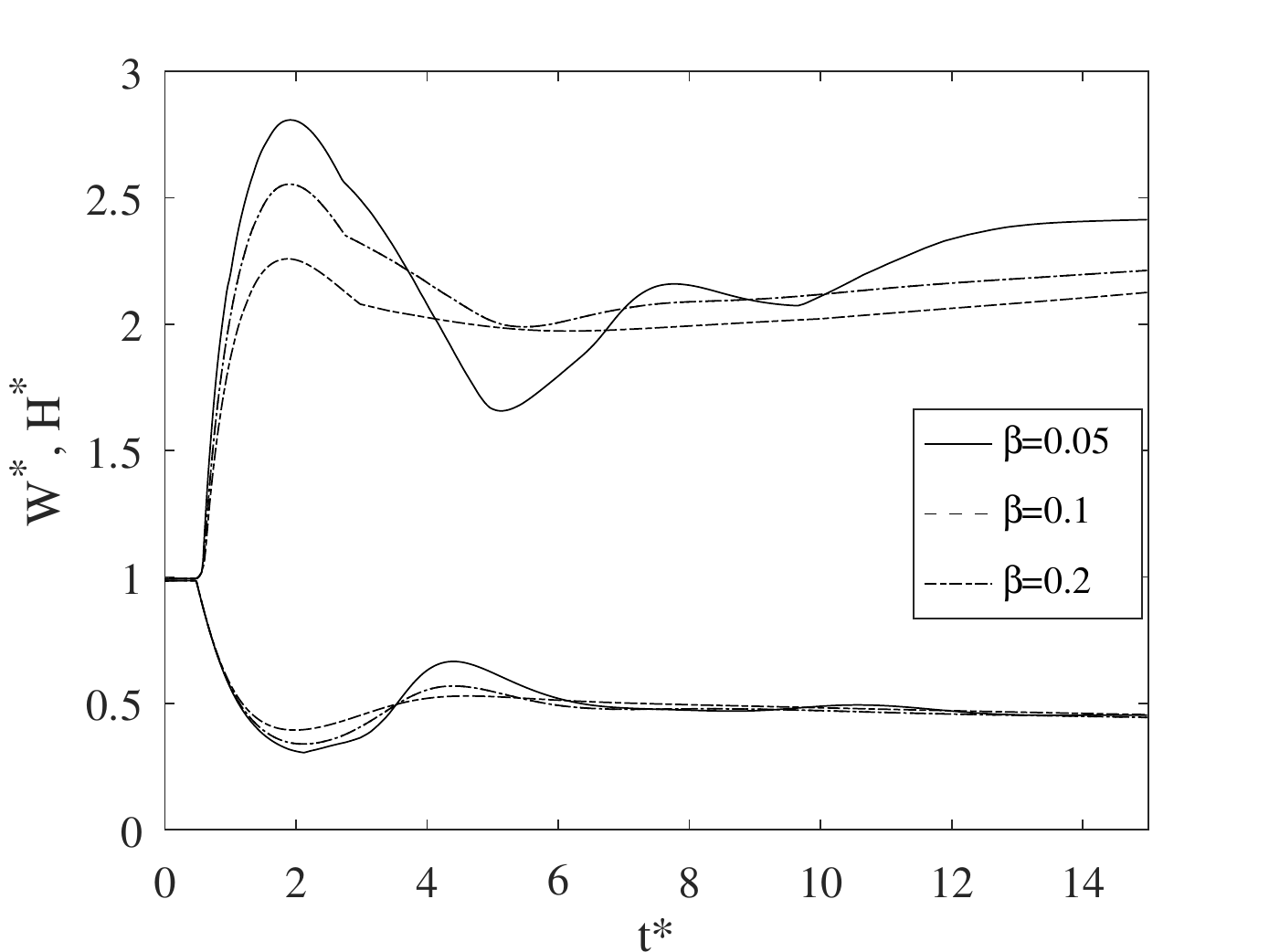}  
\caption{Variation of non-dimensional width $W^{\star}$ and height $H^{\star}$ of Oldroyd B drops with non-dimensional time $t^{\star}$. The left panel shows the effects of increasing $De$ with $\beta=0.1$. The right panel shows the effect of decreasing $\beta$, with $De=5$. In all cases, the resolution is $W/\delta{r}=120$.\label{fig:drops_ob_var}}
\end{figure}

Figure~\ref{fig:drops_ob_varde} shows a drop of Oldroyd B fluid at a number of instants in time, as $De$ is increased. The left column shows the Newtonian limit with $\lambda=De=0$, and we see a simple spreading of the drop, with no rebound (by which we mean retraction of the spreading, not detachment from the solid plate). With increasing $De$ the maximum drop width increases (visible in the fourth row, corresponding to $t^{\star}=2.16$), with larger values of $tr\left(\bm{A}\right)$ within the drop - corresponding to a greater degree of polymer stretching. The viscoelastic drops rebound, with the time to rebound proportional to $De$, and the polymer deformation reduces as the stored elastic energy is released. The left panel of Figure~\ref{fig:drops_ob_var} shows the variation of $W^{\star}$ and $H^{\star}$ with $t^{\star}$ for the Oldroyd B fluid as $De$ is increased (with $\beta=0.1$). For non-zero $De$, the drops rebound, and the degree of rebound is greater for smaller $De$, whilst for moderate $De\le100$, the time after impact at which the maximum value of $W^{\star}$ occurs is proportional to $De$. For large $De$ there is no rebound of the drop, which spreads continuously, as the elastic time-scale is long compared with the time-scale over which the spreading motion occurs. As $De$ is increased beyond $500$, the solution breaks down at the time of drop impact. The right panel of Figure~\ref{fig:drops_ob_var} shows the variation of $W^{\star}$ and $H^{\star}$ with $t^{\star}$ as the solvent viscosity ratio $\beta$ is reduced (with $De=5$). For all three values of $\beta$ we see peak $W^{\star}$ occur at the same time, but with larger values for smaller $\beta$, as the drop spreads more with lower solvent viscosity. The degree of rebound is significantly affected by $\beta$; drops with lower solvent viscosity spread more, and hence store more elastic energy, before rebounding more strongly. 

\begin{figure}
\includegraphics[width=0.99\textwidth]{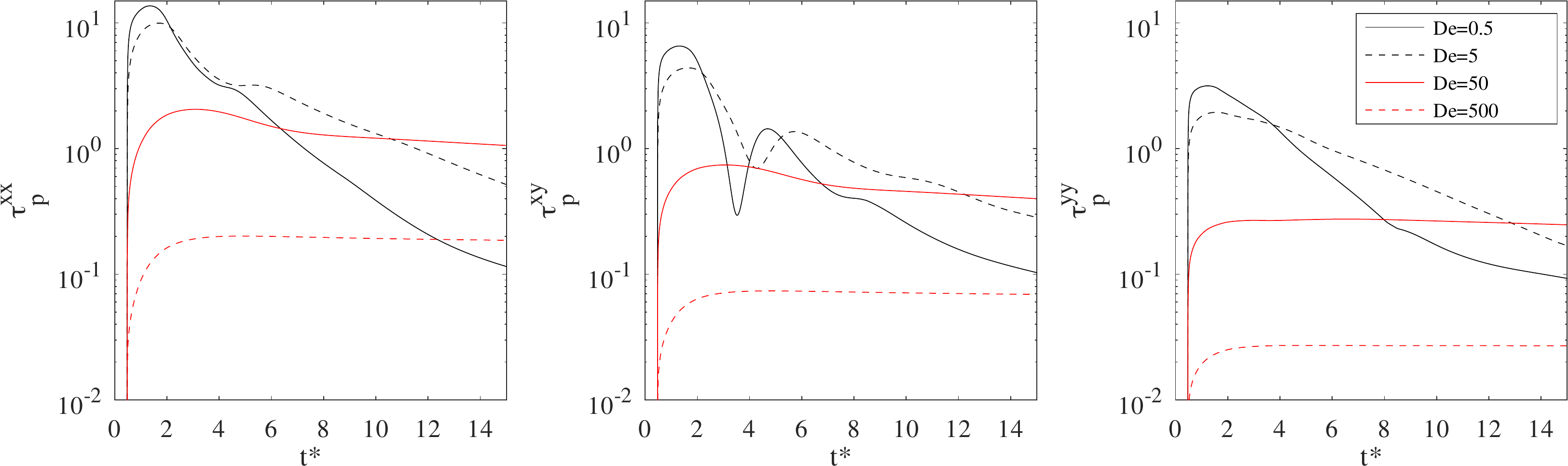}
\caption{Variation of the components of polymeric stress $\tau_{p}^{xx}$ (left) $\tau_{p}^{xy}$ (centre) and $\tau_{p}^{yy}$ (right) at a location $\left(x,y\right)=\left(0,W/8\right)$, with dimensionless time $t^{\star}$, for several values of $De$, with $Re=4$ and $\beta=0.1$. The polymeric stress components have been non-dimensionalised with a characteristic solvent stress $\eta_{s}U_{d}/W$.\label{fig:dropstress}} 
\end{figure}

As $\beta$ approaches zero, the thickness of the viscous boundary layer at the instant of impact also tends towards zero. In the absence of a scheme with adaptive resolution, it is not feasible to resolve this boundary layer, resulting in a limit to how far $\beta$ can be reduced whilst the simulation remains stable. In the present case (with $Re=4$) we find this limit to be $\beta=0.05$. Decreasing the Reynolds number has the effect of allowing larger $De$ and smaller $\beta$ to be simulated with a given resolution, as the spreading velocity is reduced, and the thickness of the viscous boundary layer is increased. Figure~\ref{fig:dropstress} shows the time variation of the polymeric stress components at a distance $W/8$ above the plate (at $\left(x,y\right)=\left(0,W/8\right)$), for Deborah numbers from $De=0.5$ to $De=500$ (with $Re=4$ and $\beta=0.1$). In all components of the stress, we see increasing $De$ reduces the peak stress during the early stages of spreading. For $De=50$ and $De=500$, the decay in stress after the peak is extremely slow, whilst for $De=0.5$ and $De=5$, the stress decays much more quickly, with a clear secondary peak in the shear component $\tau_{p}^{xy}$ corresponding to the rebound of the drop visible in Figure~\ref{fig:drops_ob_varde}.

\begin{figure}
\includegraphics[width=0.49\textwidth]{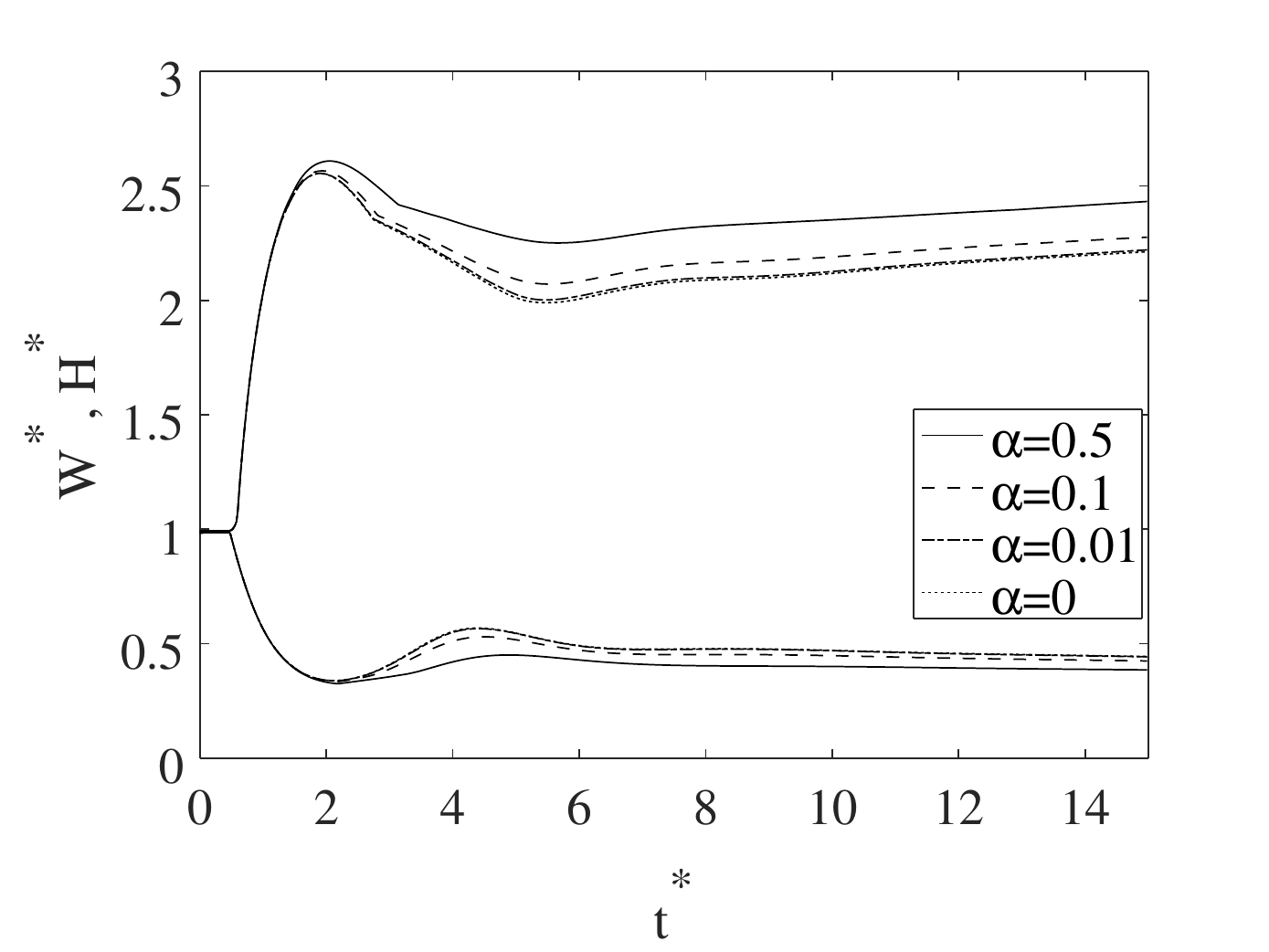}
\includegraphics[width=0.49\textwidth]{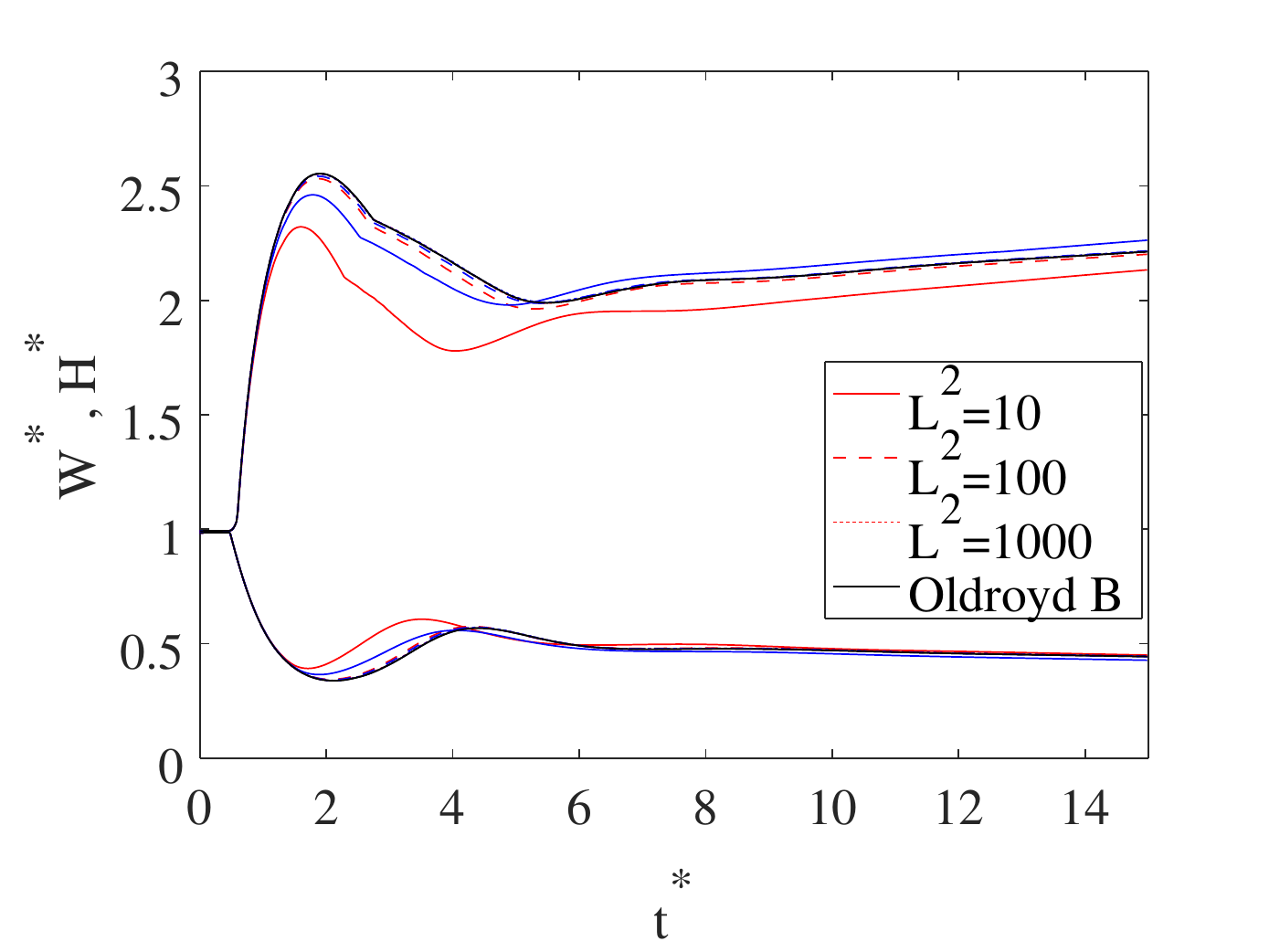}  
\caption{Variation of non-dimensional drop width $W^{\star}$ and height $H^{\star}$ with non-dimensional time $t^{\star}$, for $De=5$, $\beta=0.1$, and $W/\delta{r}=120$. The left panel shows the effects of increasing non-linearity $\alpha$ in a Giesekus fluid. The right panel shows the effect of increasing non-linearity (reducing $L^{2}$) in the FENE fluids. Blue lines correspond to the FENE-P model, and red lines correspond to the FENE-CR model. The black line is an Oldroyd B fluid.\label{fig:drops_fene_g}}
\end{figure}

Finally for this test problem we simulate drop impacts with non-linear constitutive models. The left panel of Figure~\ref{fig:drops_fene_g} shows the variation of $W^{\star}$ and $H^{\star}$ with $t^{\star}$ for a Giesekus drop with various values of mobility parameter $\alpha$. For larger $\alpha$, the drop rebounds less, and spreads more, as the fluid exhibits a greater degree of shear thinning. The right panel of Figure~\ref{fig:drops_fene_g} shows the variation of $W^{\star}$ and $H^{\star}$ with $t^{\star}$ for a FENE-P (blue lines) and FENE-CR (red lines) drops, for different values of $L^{2}$. With smaller $L^{2}$ the drop rebounds earlier and more sharply. The FENE-P model predicts a slightly greater degree of spreading at late times and than FENE-CR and Oldroyd B models, which is consistent with the shear thinning behaviour introduced by the Peterlin closure approximation.

\begin{figure}
\includegraphics[width=0.9\textwidth]{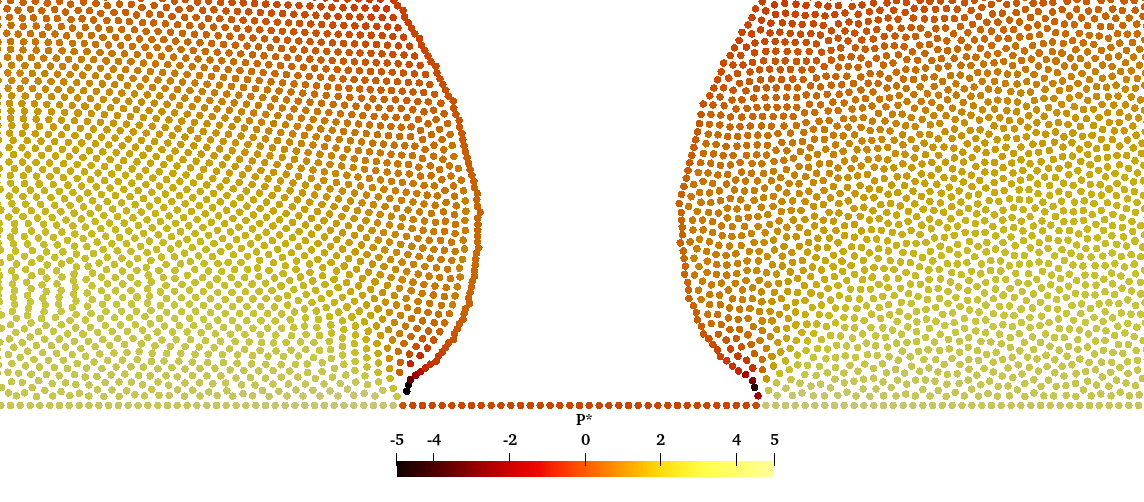}
\caption{The pressure field and particle distribution in Newtonian drops at $t^{\star}=0.72$, without (left) and with (right) the final term in~\eqref{eq:gradC}. \label{fig:shift}}
\end{figure}

Figure~\ref{fig:shift} shows a close up of the non-dimensional pressure field $P^{\star}=pW/U_{d}\eta_{0}$ at time $t^{\star}=0.72$ for a Newtonian drop. The left image is with the omission of the final term in~\eqref{eq:gradC}, whilst the term is included in right image. The clustering of particles near free surfaces is a common feature of free surface flows in SPH, and can be clearly seen in the left image of Figure~\ref{fig:shift}. This effect is significantly reduced in the right image. The final term in~\eqref{eq:gradC} has the effect of applying a repulsive force to all internal particles, causing the internal particles to shift \emph{relative to} the free surface particles such that the clustering of particles at the free surface is reduced. This term has the added benefit of reducing anisotropy in the particle distribution away from the free surface, as can be seen by comparing the far left and far right of Figure~\ref{fig:shift}. 

\subsection{Buckling jets}

Finally we use our numerical method to simulate buckling viscoelastic jets. A jet of width $D=0.006m$ flows with an initial velocity $U_{0}=0.2ms^{-1}$ and zero initial stress from an opening a height of $H=12D$ above a solid surface. The jet is subject to gravity $g=9.81ms^{-2}$, and has total viscosity $\eta_{0}=8.8Pa\cdot{s}$, and density $\rho=1.023\times10^{3}kgm^{-3}$, yielding a Reynolds number of $Re=0.14$. These properties approximately match those of a typical commericially available shampoo, as used in the experimental work of~\cite{king_2019}. The fluid has a viscosity ratio of $\beta=0.2$. The resolution is $\delta{r}=D/20$, and we omit the EVSS scheme, obtaining stable results with $\alpha_{V}=0$. We vary the constitutive model, non-linearity parameters, and orifice Weissenberg number, which is defined as $Wi=\lambda{U}_{0}/D$. 

\begin{figure}
\includegraphics[width=0.99\textwidth]{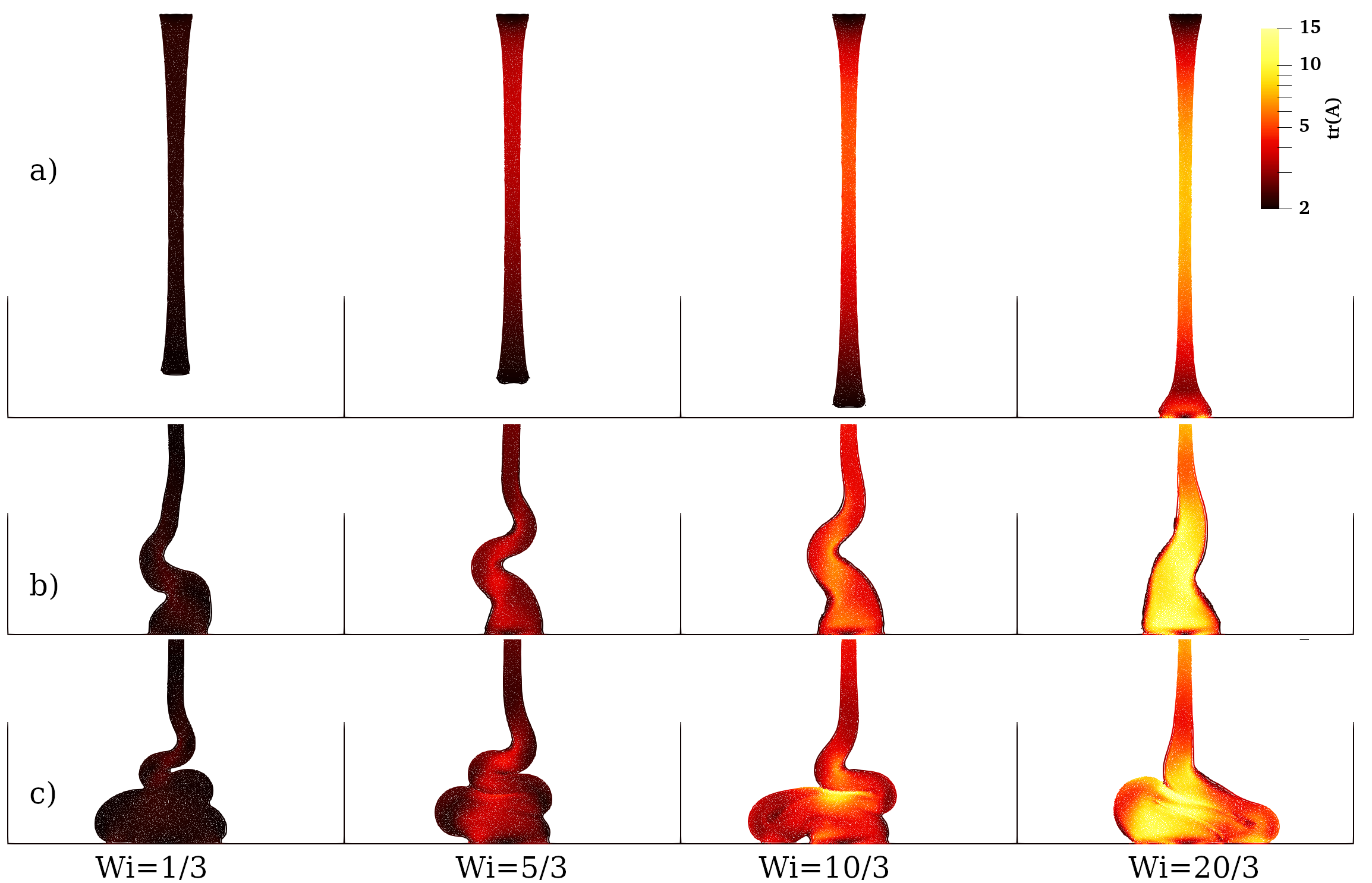}
\caption{Buckling jets of an Oldroyd B fluid at various times, with $\beta=0.2$ and for various $Wi$. From left to right, the columns correspond to $Wi=1/3$, $Wi=5/3$, $Wi=10/3$, and $Wi=20/3$. Each row corresponds to a time: a) $t=0.18s$, b) $t=0.27s$, and c) $t=0.35$. Colour indicates the trace of the conformation tensor.\label{fig:jets_ob}}
\end{figure}

Figure~\ref{fig:jets_ob} shows jets of an Oldroyd B fluid at several times, for a range of $Wi$. The colour indicates the trace of the conformation tensor, and it is clear (as expected) that for larger $Wi$, values of $tr\left(\bm{A}\right)$ throughout the jet are generally larger, indicating greater polymeric deformation. In row a) ($t=0.18s$) the difference in fall speeds is apparent, as a higher $Wi$ results in a jet which undergoes a greater degree of thinning as it falls, and impacts on the base of the container earlier. In rows b) and c) we see that for larger $Wi$ the jet spreads more before buckling, and buckles later. For lower $Wi$ the buckling is more regular, whilst for higher $Wi$ the buckling is irregular, as it interacts with the slumping of the heap.

\begin{figure}
\includegraphics[width=0.99\textwidth]{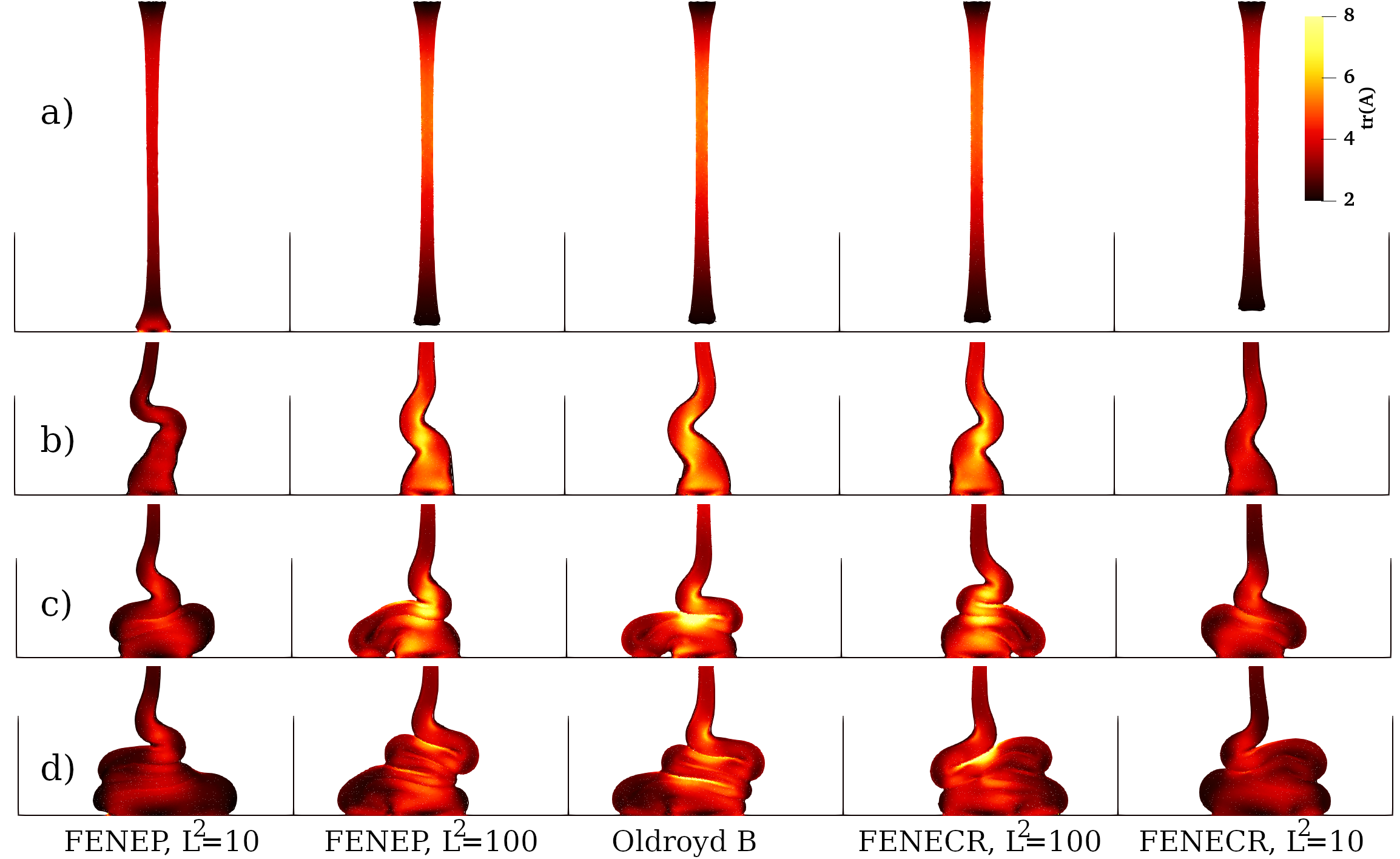}
\caption{Buckling jets of several viscoelastic fluids at various times, with $Wi=10/3$ and $\beta=0.2$. From left to right, the columns correspond to FENE-P with $L^{2}=10$ and $L^{2}=100$, Oldroyd B, and FENE-CR with $L^{2}=100$ and $L^{2}=10$. Each row corresponds to a time: a) $t=0.18s$, b) $t=0.27s$, c) $t=0.35$, and d) $t=0.43s$. Colour indicates the trace of the conformation tensor.\label{fig:jets_fene}}
\end{figure}

Figure~\ref{fig:jets_fene} shows jets of a FENE-P (left two columns), FENE-CR (right-two columns) and an Oldroyd B (centre column) fluid, at several times, with $Wi=10/3$. In row a) we see that the non-linearity in the FENE-P model results the jet falling faster (the impact on the container base is earlier for $L^{2}=10$ than for $L^{2}=100$), with increased stretching, without an increase (relative to the Oldroyd B fluid) in the molecular deformation. This is consistent with the shear-thinning properties of the FENE-P model. Conversely, the non-linearity in the FENE-CR model results the jet falling more slowly, and a reduction in stretching. In row b) it is clear that the constitutive model influences the buckling, with the Oldroyd B and FENE-P fluids initially buckling to the left, whilst the FENE-CR model initially buckles to the right. At later times (e.g. row d)) we see that a greater degree of non-linearity (i.e. smaller $L^{2}$) limits the value of the conformation tensor trace and the degree of polymeric deformation. Finally, we observed that for both the FENE-P and the FENE-CR models, with a smaller value of $L^{2}$ the jet buckles at a slightly higher frequency and with a smaller amplitude.

\begin{figure}
\includegraphics[width=0.99\textwidth]{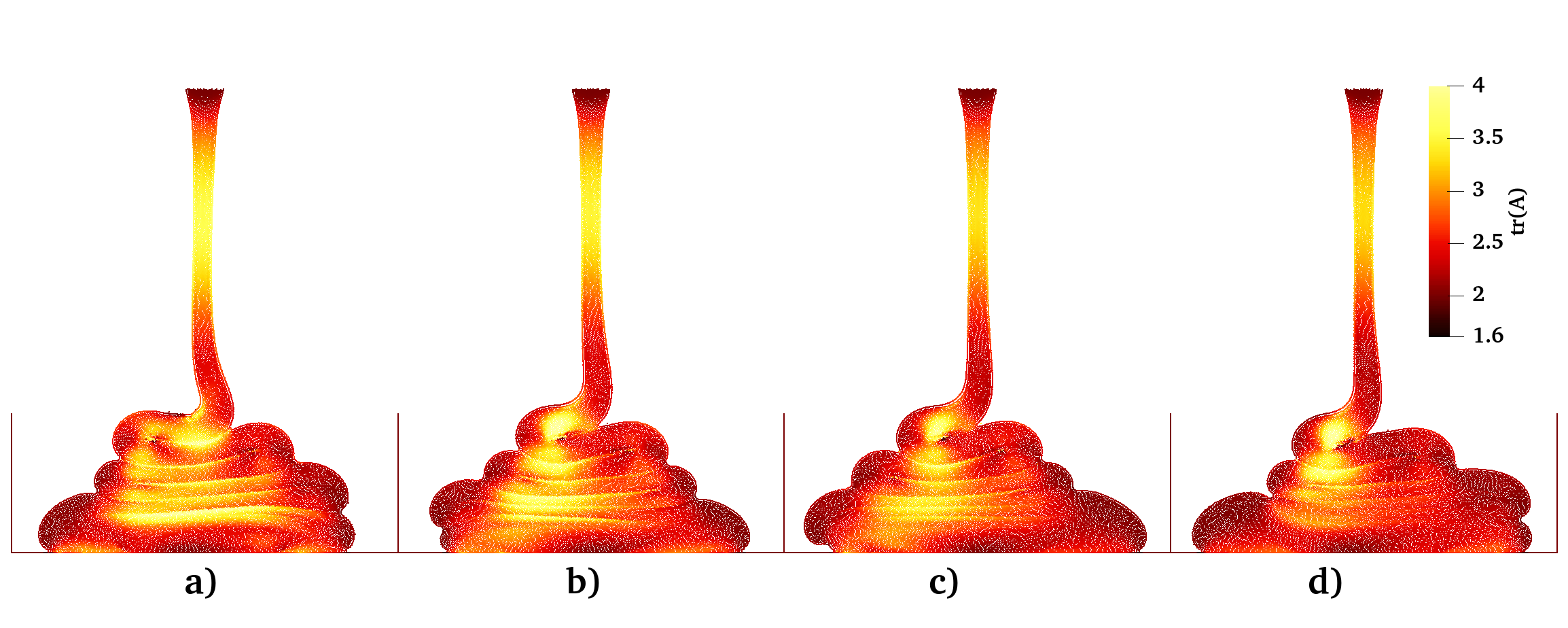}
\caption{Buckling jets of a Giesekus fluid at time $t=0.7s$, with $Wi=10/3$ and $\beta=0.2$, for various values of nonlinearity parameter $\alpha$: a) $\alpha=0$, b) $\alpha=0.01$, c) $\alpha=0.1$, and d) $\alpha=0.2$.\label{fig:jets_g}}
\end{figure}

Finally Figure~\ref{fig:jets_g} shows jets of a Giesekus fluid for time $t=0.7s$ for various values of mobility parameter $\alpha$. Again, the colour indicates $tr\left(\bm{A}\right)$.  With increasing $\alpha$ the coiling frequency is increased, with greater (transverse) amplitude, and the heap subsides more quickly due to the increased shear thinning. With each fold, a shear layer is formed, and within the heap there is polymer extension as the folds of the jet undergo transverse stretching whilst slumping, driven by the younger folds above. This extension is larger for smaller $\alpha$. \citet{tome_2019} also consider the simulation of Giesekus jets, and although the parameters used in~\cite{tome_2019} differ in detail, the behaviour we see is qualitatively similar.

\section{Conclusions}\label{conc}

Smoothed Particle Hydrodynamics (SPH) has significant potential for simulations of viscoelastic flows, particularly those involving free surfaces, but advances in modelling viscoelastic fluids in SPH to date have lagged behind other methods. In this work we have presented a robust incompressible SPH formulation capable of simulating a broad range of viscoelastic flows. This scheme presents an advance in the state-of-the-art in SPH, greatly increasing the robustness of the method at high Weissenberg numbers. The key strengths of the method are as follows:
\begin{enumerate}
\item By introducing the log-conformation tensor formulation for the first time to SPH, we ensure that the viscoelastic stresses in our simulations correspond to physically meaningful molecular deformations. More importantly, the High Weissenberg Number Problem (HWNP) is avoided, enabling simulations of at significantly higher Weissenberg numbers than previously possible with SPH.
\item By introducing an elasto-viscous stress splitting (EVSS) scheme to SPH, we are able to simulate flows with very small or zero solvent viscosity.
\item We tailor the widely used Fickian shifting procedure to low Reynolds number flows, which results in improved particle distributions at free surfaces. We also introduce a free-surface pressure boundary condition which better matches the physics than in other incompressible SPH schemes.
\item By formulating the consitutive models in terms of strain and relaxation functions, a wide variety of constitutive models may be easily implemented, without making structural changes to the method.
\end{enumerate}
We have demonstrated the ability of our method to simulate flows at high Weissenberg numbers (up to $Wi=85$) for a range of internal and free surface problems. Numerical results compare well with analytical solutions and published data. In the tests presented here we found the limit of accuracy and stability of the method appears to be related to an interaction between the shifting procedure (required for stability), and the polymeric stress. Exploring this issue is an active area of research for the authors. Further developments which are planned include the addition of a surface tension model, which is known to be challenging in an incompressible SPH framework, and the development of a multiphase scheme. 

\section*{Acknowledgements}
We are grateful for financial support from the Leverhulme Trust, via Research Project Grants RPG-2017-144 and RPG-2019-206. We thank two anonymous reviewers for some insightful comments which have helped improve this work.

\bibliographystyle{elsarticle-num-names}
\bibliography{jrckbib}

\end{document}